\documentclass[trackchanges,resetfootnote,twocolumn]{aastex7}
\usepackage{CJK}

\shorttitle{Pair Search for Strong Lenses in DESI DR1}
\shortauthors{Y.-M. Hsu et al.}
\submitjournal{ApJS}

\begin{document}
\begin{CJK*}{UTF8}{bsmi}

\title{A New Way to Discover Strong Gravitational Lenses: Pair-wise Spectroscopic Search from DESI DR1}

\author[orcid=0000-0002-6876-8492,gname='Yuan-Ming',sname=Hsu]{Yuan-Ming Hsu (許淵明)}
\affiliation{Department of Physics, National Taiwan University, No. 1, Section 4, Roosevelt Road, Taipei 106319, Taiwan}
\email{noreply01@example.edu}

\author[orcid=0000-0001-8156-0330,gname=Xiaosheng,sname=Huang]{Xiaosheng Huang}
\affiliation{Department of Physics \& Astronomy, University of San Francisco, 2130 Fulton Street, San Francisco, CA 94117, USA}
\affiliation{Lawrence Berkeley National Laboratory, 1 Cyclotron Road, Berkeley, CA 94720, USA}
\email[show]{xhuang22@usfca.edu}

\author[orcid=0000-0002-0385-0014,gname=Christopher,sname=Storfer]{Christopher J. Storfer}
\affiliation{Institute for Astronomy, University of Hawai`i, 2680 Woodlawn Drive, Honolulu, HI 96822, USA}
\affiliation{Lawrence Berkeley National Laboratory, 1 Cyclotron Road, Berkeley, CA 94720, USA}
\email{noreply03@example.edu}

\author[orcid=0009-0009-8667-763X,gname='Jose Carlos',sname=Inchausti]{Jose Carlos Inchausti}
\affiliation{Department of Physics \& Astronomy, University of San Francisco, 2130 Fulton Street, San Francisco, CA 94117, USA}
\email{noreply04@example.edu}

\author[orcid=0000-0002-5042-5088,gname=David,sname=Schlegel]{David Schlegel}
\affiliation{Lawrence Berkeley National Laboratory, 1 Cyclotron Road, Berkeley, CA 94720, USA}
\email{noreply05@example.edu}

\author[orcid=0000-0002-2733-4559,gname=John,sname=Moustakas]{John Moustakas}
\affiliation{Department of Physics and Astronomy, Siena College, 515 Loudon Road, Loudonville, NY 12211, USA}
\email{noreply06@example.edu}

\author[gname='Jessica Nicole',sname='Aguilar']{J.~Aguilar}
\affiliation{Lawrence Berkeley National Laboratory, 1 Cyclotron Road, Berkeley, CA 94720, USA}
\email{noreply07@example.edu}

\author[orcid=0000-0001-6098-7247,gname='Steven',sname='Ahlen']{S.~Ahlen}
\affiliation{Department of Physics, Boston University, 590 Commonwealth Avenue, Boston, MA 02215 USA}
\email{noreply08@example.edu}

\author[orcid=0000-0003-2923-1585,gname='Abhijeet',sname='Anand']{A.~Anand}
\affiliation{Lawrence Berkeley National Laboratory, 1 Cyclotron Road, Berkeley, CA 94720, USA}
\email{noreply09@example.edu}

\author[orcid=0000-0003-4162-6619,gname='Stephen',sname='Bailey']{S.~Bailey}
\affiliation{Lawrence Berkeley National Laboratory, 1 Cyclotron Road, Berkeley, CA 94720, USA}
\email{noreply10@example.edu}

\author[orcid=0000-0001-9712-0006,gname='Davide',sname='Bianchi']{D.~Bianchi}
\affiliation{Dipartimento di Fisica ``Aldo Pontremoli'', Universit\`a degli Studi di Milano, Via Celoria 16, I-20133 Milano, Italy}
\affiliation{INAF-Osservatorio Astronomico di Brera, Via Brera 28, 20122 Milano, Italy}
\email{noreply11@example.edu}

\author[gname='David',sname='Brooks']{D.~Brooks}
\affiliation{Department of Physics \& Astronomy, University College London, Gower Street, London, WC1E 6BT, UK}
\email{noreply12@example.edu}

\author[orcid=0000-0001-7316-4573,gname='Francisco Javier',sname='Castander']{F.~J.~Castander}
\affiliation{Institut d'Estudis Espacials de Catalunya (IEEC), c/ Esteve Terradas 1, Edifici RDIT, Campus PMT-UPC, 08860 Castelldefels, Spain}
\affiliation{Institute of Space Sciences, ICE-CSIC, Campus UAB, Carrer de Can Magrans s/n, 08913 Bellaterra, Barcelona, Spain}
\email{noreply13@example.edu}

\author[gname='Todd',sname='Claybaugh']{T.~Claybaugh}
\affiliation{Lawrence Berkeley National Laboratory, 1 Cyclotron Road, Berkeley, CA 94720, USA}
\email{noreply14@example.edu}

\author[orcid=0000-0002-2169-0595,gname='Andrei',sname='Cuceu']{A.~Cuceu}
\affiliation{Lawrence Berkeley National Laboratory, 1 Cyclotron Road, Berkeley, CA 94720, USA}
\email{noreply15@example.edu}

\author[orcid=0000-0002-1769-1640,gname='Axel',sname='de la Macorra']{A.~de la Macorra}
\affiliation{Instituto de F\'{\i}sica, Universidad Nacional Aut\'{o}noma de M\'{e}xico,  Circuito de la Investigaci\'{o}n Cient\'{\i}fica, Ciudad Universitaria, Cd. de M\'{e}xico  C.~P.~04510,  M\'{e}xico}
\email{noreply16@example.edu}

\author[orcid=0000-0003-0928-2000,gname='John',sname='Della Costa']{J.~Della~Costa}
\affiliation{Department of Astronomy, San Diego State University, 5500 Campanile Drive, San Diego, CA 92182, USA}
\affiliation{NSF NOIRLab, 950 N. Cherry Ave., Tucson, AZ 85719, USA}
\email{noreply17@example.edu}

\author[orcid=0000-0002-4928-4003,gname='Arjun',sname='Dey']{Arjun~Dey}
\affiliation{NSF NOIRLab, 950 N. Cherry Ave., Tucson, AZ 85719, USA}
\email{noreply18@example.edu}

\author[orcid=0000-0002-5665-7912,gname='Biprateep',sname='Dey']{Biprateep~Dey}
\affiliation{Department of Astronomy \& Astrophysics, University of Toronto, Toronto, ON M5S 3H4, Canada}
\affiliation{Department of Physics \& Astronomy and Pittsburgh Particle Physics, Astrophysics, and Cosmology Center (PITT PACC), University of Pittsburgh, 3941 O'Hara Street, Pittsburgh, PA 15260, USA}
\email{noreply19@example.edu}

\author[gname='Peter',sname='Doel']{P.~Doel}
\affiliation{Department of Physics \& Astronomy, University College London, Gower Street, London, WC1E 6BT, UK}
\email{noreply20@example.edu}

\author[orcid=0000-0002-2890-3725,gname='Jaime E.',sname='Forero-Romero']{J.~E.~Forero-Romero}
\affiliation{Departamento de F\'isica, Universidad de los Andes, Cra. 1 No. 18A-10, Edificio Ip, CP 111711, Bogot\'a, Colombia}
\affiliation{Observatorio Astron\'omico, Universidad de los Andes, Cra. 1 No. 18A-10, Edificio H, CP 111711 Bogot\'a, Colombia}
\email{noreply21@example.edu}

\author[orcid=0000-0001-9632-0815,gname='Enrique',sname='Gaztañaga']{E.~Gaztañaga}
\affiliation{Institut d'Estudis Espacials de Catalunya (IEEC), c/ Esteve Terradas 1, Edifici RDIT, Campus PMT-UPC, 08860 Castelldefels, Spain}
\affiliation{Institute of Cosmology and Gravitation, University of Portsmouth, Dennis Sciama Building, Portsmouth, PO1 3FX, UK}
\affiliation{Institute of Space Sciences, ICE-CSIC, Campus UAB, Carrer de Can Magrans s/n, 08913 Bellaterra, Barcelona, Spain}
\email{noreply22@example.edu}

\author[orcid=0000-0003-3142-233X,gname='Satya',sname='Gontcho A Gontcho']{S.~Gontcho A Gontcho}
\affiliation{Lawrence Berkeley National Laboratory, 1 Cyclotron Road, Berkeley, CA 94720, USA}
\affiliation{University of Virginia, Department of Astronomy, Charlottesville, VA 22904, USA}
\email{noreply23@example.edu}

\author[gname='Gaston',sname='Gutierrez']{G.~Gutierrez}
\affiliation{Fermi National Accelerator Laboratory, PO Box 500, Batavia, IL 60510, USA}
\email{noreply24@example.edu}

\author[orcid=0000-0001-6558-0112,gname='Dragan',sname='Huterer']{D.~Huterer}
\affiliation{Department of Physics, University of Michigan, 450 Church Street, Ann Arbor, MI 48109, USA}
\email{noreply25@example.edu}

\author[orcid=0000-0003-0201-5241,gname='Dick',sname='Joyce']{R.~Joyce}
\affiliation{NSF NOIRLab, 950 N. Cherry Ave., Tucson, AZ 85719, USA}
\email{noreply26@example.edu}

\author[gname='Robert',sname='Kehoe']{R.~Kehoe}
\affiliation{Department of Physics, Southern Methodist University, 3215 Daniel Avenue, Dallas, TX 75275, USA}
\email{noreply27@example.edu}

\author[orcid=0000-0002-8828-5463,gname='David',sname='Kirkby']{D.~Kirkby}
\affiliation{Department of Physics and Astronomy, University of California, Irvine, 92697, USA}
\email{noreply28@example.edu}

\author[orcid=0000-0003-3510-7134,gname='Theodore',sname='Kisner']{T.~Kisner}
\affiliation{Lawrence Berkeley National Laboratory, 1 Cyclotron Road, Berkeley, CA 94720, USA}
\email{noreply29@example.edu}

\author[orcid=0000-0001-6356-7424,gname='Anthony',sname='Kremin']{A.~Kremin}
\affiliation{Lawrence Berkeley National Laboratory, 1 Cyclotron Road, Berkeley, CA 94720, USA}
\email{noreply30@example.edu}

\author[gname='Ofer',sname='Lahav']{O.~Lahav}
\affiliation{Department of Physics \& Astronomy, University College London, Gower Street, London, WC1E 6BT, UK}
\email{noreply31@example.edu}

\author[orcid=0000-0003-1838-8528,gname='Martin',sname='Landriau']{M.~Landriau}
\affiliation{Lawrence Berkeley National Laboratory, 1 Cyclotron Road, Berkeley, CA 94720, USA}
\email{noreply32@example.edu}

\author[orcid=0000-0001-7178-8868,gname='Laurent',sname='Le Guillou']{L.~Le~Guillou}
\affiliation{Sorbonne Universit\'{e}, CNRS/IN2P3, Laboratoire de Physique Nucl\'{e}aire et de Hautes Energies (LPNHE), FR-75005 Paris, France}
\email{noreply33@example.edu}

\author[orcid=0000-0003-4962-8934,gname='Marc',sname='Manera']{M.~Manera}
\affiliation{Departament de F\'{i}sica, Serra H\'{u}nter, Universitat Aut\`{o}noma de Barcelona, 08193 Bellaterra (Barcelona), Spain}
\affiliation{Institut de F\'{i}sica d’Altes Energies (IFAE), The Barcelona Institute of Science and Technology, Edifici Cn, Campus UAB, 08193, Bellaterra (Barcelona), Spain}
\email{noreply34@example.edu}

\author[orcid=0000-0002-1125-7384,gname='Aaron',sname='Meisner']{A.~Meisner}
\affiliation{NSF NOIRLab, 950 N. Cherry Ave., Tucson, AZ 85719, USA}
\email{noreply35@example.edu}

\author[gname='Ramon',sname='Miquel']{R.~Miquel}
\affiliation{Instituci\'{o} Catalana de Recerca i Estudis Avan\c{c}ats, Passeig de Llu\'{\i}s Companys, 23, 08010 Barcelona, Spain}
\affiliation{Institut de F\'{i}sica d’Altes Energies (IFAE), The Barcelona Institute of Science and Technology, Edifici Cn, Campus UAB, 08193, Bellaterra (Barcelona), Spain}
\email{noreply36@example.edu}

\author[orcid=0000-0001-9070-3102,gname='Seshadri',sname='Nadathur']{S.~Nadathur}
\affiliation{Institute of Cosmology and Gravitation, University of Portsmouth, Dennis Sciama Building, Portsmouth, PO1 3FX, UK}
\email{noreply37@example.edu}

\author[orcid=0000-0003-3188-784X,gname='Nathalie',sname='Palanque-Delabrouille']{N.~Palanque-Delabrouille}
\affiliation{IRFU, CEA, Universit\'{e} Paris-Saclay, F-91191 Gif-sur-Yvette, France}
\affiliation{Lawrence Berkeley National Laboratory, 1 Cyclotron Road, Berkeley, CA 94720, USA}
\email{noreply38@example.edu}

\author[orcid=0000-0002-0644-5727,gname='Will',sname='Percival']{W.~J.~Percival}
\affiliation{Department of Physics and Astronomy, University of Waterloo, 200 University Ave W, Waterloo, ON N2L 3G1, Canada}
\affiliation{Perimeter Institute for Theoretical Physics, 31 Caroline St. North, Waterloo, ON N2L 2Y5, Canada}
\affiliation{Waterloo Centre for Astrophysics, University of Waterloo, 200 University Ave W, Waterloo, ON N2L 3G1, Canada}
\email{noreply39@example.edu}

\author[orcid=0000-0001-7145-8674,gname='Francisco',sname='Prada']{F.~Prada}
\affiliation{Instituto de Astrof\'{\i}sica de Andaluc\'{\i}a (CSIC), Glorieta de la Astronom\'{\i}a, s/n, E-18008 Granada, Spain}
\email{noreply40@example.edu}

\author[orcid=0000-0001-6979-0125,gname='Ignasi',sname='Pérez-Ràfols']{I.~P\'erez-R\`afols}
\affiliation{Departament de F\'isica, EEBE, Universitat Polit\`ecnica de Catalunya, c/Eduard Maristany 10, 08930 Barcelona, Spain}
\email{noreply41@example.edu}

\author[gname='Graziano',sname='Rossi']{G.~Rossi}
\affiliation{Department of Physics and Astronomy, Sejong University, 209 Neungdong-ro, Gwangjin-gu, Seoul 05006, Republic of Korea}
\email{noreply42@example.edu}

\author[orcid=0000-0002-9646-8198,gname='Eusebio',sname='Sanchez']{E.~Sanchez}
\affiliation{CIEMAT, Avenida Complutense 40, E-28040 Madrid, Spain}
\email{noreply43@example.edu}

\author[gname='Michael',sname='Schubnell']{M.~Schubnell}
\affiliation{Department of Physics, University of Michigan, 450 Church Street, Ann Arbor, MI 48109, USA}
\email{noreply44@example.edu}

\author[orcid=0000-0002-3461-0320,gname='Joseph Harry',sname='Silber']{J.~Silber}
\affiliation{Lawrence Berkeley National Laboratory, 1 Cyclotron Road, Berkeley, CA 94720, USA}
\email{noreply45@example.edu}

\author[gname='David',sname='Sprayberry']{D.~Sprayberry}
\affiliation{NSF NOIRLab, 950 N. Cherry Ave., Tucson, AZ 85719, USA}
\email{noreply46@example.edu}

\author[orcid=0000-0003-1704-0781,gname='Gregory',sname='Tarlé']{G.~Tarl\'{e}}
\affiliation{Department of Physics, University of Michigan, 450 Church Street, Ann Arbor, MI 48109, USA}
\email{noreply47@example.edu}

\author[gname='Benjamin Alan',sname='Weaver']{B.~A.~Weaver}
\affiliation{NSF NOIRLab, 950 N. Cherry Ave., Tucson, AZ 85719, USA}
\email{noreply48@example.edu}

\author[orcid=0000-0001-5381-4372,gname='Rongpu',sname='Zhou']{R.~Zhou}
\affiliation{Lawrence Berkeley National Laboratory, 1 Cyclotron Road, Berkeley, CA 94720, USA}
\email{noreply49@example.edu}

\author[orcid=0000-0002-6684-3997,gname='Hu',sname='Zou']{H.~Zou}
\affiliation{National Astronomical Observatories, Chinese Academy of Sciences, A20 Datun Road, Chaoyang District, Beijing, 100101, P.~R.~China}
\email{noreply50@example.edu}

\correspondingauthor{Yuan-Ming Hsu, Xiaosheng Huang}
\collaboration{all}{DESI Collaboration}

\begin{abstract}
We present a new method to search for strong gravitational lensing systems by pairing spectra that are close together on the sky in a spectroscopic survey. We visually inspect 26,621 spectra in the Dark Energy Spectroscopic Instrument (DESI) Data Release 1 that are selected in this way.
We further inspect the 11,848 images corresponding to these spectra in the DESI Legacy Imaging Surveys Data Release 10, and obtain 2046 conventional strong gravitational lens candidates, of which 1906 are new. This constitutes the largest sample of lens candidates identified to date in spectroscopic data.
Besides the conventional candidates, we identify a new class of systems that we term ``dimple lenses''. These systems have a low-mass foreground galaxy as a lens, typically smaller in angular extent and fainter compared with the lensed background source galaxy, producing subtle surface brightness indentations in the latter. We report the discovery of 318 of these ``dimple-lens'' candidates. We suspect that these represent dwarf galaxy lensing. With follow-up observations, they could offer a new avenue to test the cold dark matter model by probing their mass profiles, stellar mass-halo mass relation, and halo mass function for $M_{\textrm{Halo}} \lesssim 10^{13}\,M_\odot$.
Thus, in total, we report 2164 new lens candidates. Our method demonstrates the power of pairwise spectroscopic analysis and provides a pathway complementary to imaging-based and single-spectrum lens searches.
\end{abstract}


\section{Introduction} \label{sec:intro}
Strong gravitational lensing occurs when a massive foreground object lies along the line of sight to a more distant background source. The gravitational field of the foreground object warps space-time and thus bends and magnifies the light from the background source, creating distinctive features such as arcs, Einstein rings, or multiple images. These configurations depend on the mass distribution of the lens and the geometry of the source, the lens, and the observer.

Strong lensing systems are powerful tools in astrophysics and cosmology. They provide a unique way to study the mass profiles, including both baryonic matter and dark matter, of galaxies and galaxy clusters \citep[e.g.][]{Newman2015}.
Lensing also enables magnified views of high-redshift source galaxies, offering insights into early galaxy formation and evolution \citep[e.g.,][]{Roberts-Borsani2022}. Furthermore, precise modeling of lens systems can constrain cosmological parameters, such as the Hubble constant, through time-delay measurements between multiple images of variable sources \citep[e.g.,][]{TDCOSMOCollaboration2025} and can also test alternative dark matter models beyond the standard cold dark matter (CDM), such as the self-interacting dark matter model \citep[e.g.,][]{Li2025}, wave dark matter model \citep[e.g.,][]{Chan2020a, Amruth2023}, and warm dark matter model \citep[e.g.,][]{Inoue2024}.

There has also been a growing interest in using strong lensing systems to test the CDM model with low-mass dark matter halos.
At cosmological distances, the focus tends to be on ``dark'' halos that have mass $M_{\textrm{Halo}} \lesssim 10^9\,M_\odot$ and do not host sufficient amounts of baryons to form stars \citep[e.g.,][]{Vegetti2010}.
However, CDM has not been sufficiently tested in the range between $10^9\text{--}10^{12}\,M_\odot$.
For example, Figure~11 in \citet{Driver2022} shows the halo mass function measured down to $M_{\textrm{Halo}} \sim 10^{13}\,M_\odot$.
In addition, they show that more than half of the dark matter mass in the universe is stored in halos with $M_{\textrm{Halo}} \lesssim 10^{13}\,M_\odot$, as CDM predicts an abundance of them \citep[e.g.,][]{Jenkins2001, Warren2006, Springel2008}.
The halos with $M_{\textrm{Halo}} \in 10^9\text{--}10^{12}\,M_\odot$ typically host dwarf galaxies.
Finding them represents the first step toward testing CDM below the typical galactic scale.
However, these are very hard to find at cosmological distances.
Recently, \citet{Silver2025} show that such dwarf galaxies can be discovered at cosmological distances by applying machine learning to imaging data from HST, JWST, and Roman.
We will show in this work that it is also possible to find these low-mass galaxies in large quantities and at cosmological distances in ground-based spectroscopic data.

Traditionally, strong lensing systems were identified through direct visual inspection of wide-field imaging surveys, searching for morphological features such as arcs and rings around massive galaxies or clusters \citep[e.g.,][]{O'Donnell2022}. While effective, this method is labor-intensive and constrained by human capacity. To overcome this limitation, citizen science projects have helped address this bottleneck by engaging thousands of volunteers to visually inspect images for potential lenses \citep{Marshall2016, More2016b, Sonnenfeld2020, Garvin2022}. In parallel, automated algorithms have been developed to search for characteristic morphologies using predefined criteria \citep{Gavazzi2014}. In recent years, machine learning algorithms such as convolutional neural networks have been increasingly applied to imaging data to automate the discovery of strong lenses, achieving remarkable success on large data sets \citep{Jacobs2019, Huang2020, Canameras2020}.

In addition to imaging-based searches, another well-established approach utilizes spectroscopic data from large-scale spectroscopic sky surveys such as the Sloan Digital Sky Survey. A strong lens candidate can be identified when a single fiber spectrum exhibits features from two objects at distinct redshifts, allowing detection of lensing systems even when the corresponding imaging features are not discernible \citep{Bolton2006}. In all cases that we are aware of, these lens candidate systems consist of a massive foreground galaxy and a background emission-line source.

The recent advent of large-scale spectroscopic surveys such as the Dark Energy Spectroscopic Instrument \citep[DESI;][]{DESICollaboration2025} has dramatically expanded the available spectroscopic data set, providing millions of spectra across a wide redshift range with unprecedented coverage and depth. This wealth of spectroscopic information opens up new discovery space beyond what is accessible through lens searching using imaging data or single-fiber spectra.

In this paper, we present a new method to search for strong gravitational lens candidates by pairing spectra from a spectroscopic sky survey and visually inspecting both the spectra and corresponding imaging data. In Section~\ref{sec:obs}, we describe the observations and data used in this work. In Section~\ref{sec:spec}, we describe the procedure for selecting spectra from the data. In Section~\ref{sec:img}, we detail the visual inspection process and present the resulting list of candidates. We discuss our findings in Section~\ref{sec:discuss} and summarize in Section~\ref{sec:summary}.

\section{Observations} \label{sec:obs}
In this work, we use spectroscopic data from Data Release 1 (DR1) of the Dark Energy Spectroscopic Instrument \citep{DESICollaboration2025}, along with imaging and photometric data from Data Release 10 (DR10) of the DESI Legacy Imaging Surveys \citep[D.~J.~Schlegel in prep.]{Dey2019}.

\subsection{DESI DR1} \label{subsec:desi}
DESI is a multi-fiber spectrograph installed on the Mayall 4 m Telescope at Kitt Peak National Observatory, designed to measure redshifts of galaxies and quasars \citep{DESICollaboration2016a, DESICollaboration2022}. The instrument deploys 5000 robotically actuated fibers of diameter $1\farcs5$ across a focal plane with a diameter of 3.2 degrees \citep{DESICollaboration2016b, Silber2023, Miller2024, Poppett2024}, and DESI's observing strategy is optimized to maximize survey efficiency \citep{Schlafly2023}. The spectrographs cover a wavelength range of $3600\text{--}9800\,\text{\AA}$, split across three arms (blue, red, and near-infrared), with a resolving power ranging from $R\sim2000$ at the blue end to $R\sim5500$ at the red end \citep{DESICollaboration2022}. This resolution is sufficient to resolve key spectral features such as the [\ion{O}{2}]$\lambda\lambda3726,3729$ doublet, the [\ion{O}{3}]$\lambda\lambda4959,5007$ doublet, $\text{H}\beta$ and $\text{H}\alpha$ emission lines, as well as absorption features like Ca H\&K lines \citep[e.g.,][]{Lan2023}.

DESI DR1 provides spectra for over 18 million unique objects, processed through a sophisticated data reduction pipeline \citep{Guy2023}. Redshifts and classifications are derived using the template-fitting pipeline \texttt{Redrock} \citep[S.~Bailey et al., in prep.]{Brodzeller2023, Anand2024}, and the results are compiled into merged redshift catalogs covering all targeted sources. DR1 also includes multiple value-added catalogs, among which the \texttt{FastSpecFit} Spectral Synthesis and Emission-Line Catalog provides derived spectrophotometric quantities, such as velocity dispersions, as well as additional spectral measurements obtained from spectral fitting \citep[in prep.]{Moustakas2023}. DESI's target selection is based on photometry from the DESI Legacy Imaging Surveys \citep{Myers2023}. There are four main extragalactic target classes: the Bright Galaxy Survey \citep[BGS;][]{Hahn2023}, targeting galaxies at $z\lesssim 0.6$; Luminous Red Galaxies \citep[LRGs;][]{Zhou2023} at $0.4\lesssim z\lesssim 1.0$; Emission Line Galaxies \citep[ELGs;][]{Raichoor2023} at $0.6\lesssim z\lesssim 1.6$; and Quasars \citep[QSOs;][]{Chaussidon2023} over a broad redshift range.

\subsection{DESI Legacy Imaging Surveys DR10} \label{subsec:desi_ls}
\citet{Dey2019} describe the DESI Legacy Imaging Surveys (hereafter the DESI Legacy Surveys), which combine three imaging surveys: the Dark Energy Camera Legacy Survey (DECaLS), the Mayall z-band Legacy Survey (MzLS), and the Beijing-Arizona Sky Survey (BASS).
DECaLS has been conducted using the 4 m Blanco telescope, covering $\sim9000\deg^2$ of the sky in both the North and South Galactic Cap regions, at decl. $<32\deg$ and $<34\deg$, respectively. BASS has been carried out in the $g$ and $r$ bands using the Bok 2.3 m telescope, while MzLS has imaged the $z$ band using the 4 m Mayall telescope; both BASS and MzLS target the North Galactic Cap with $\text{decl.}>32\deg$, covering $\sim5000\deg^2$.
Together, they cover more than $14,000\deg^2$ of extragalactic sky in three optical bands ($g$, $r$, $z$), providing photometrically calibrated imaging with delivered image quality (FWHM) of $1\farcs3$ ($g$), $1\farcs2$ ($r$), and $1\farcs1$ ($z$) for DECaLS; $1\farcs6$ ($g$) and $1\farcs5$ ($r$) for BASS; and $1\farcs0$ ($z$) for MzLS. The surveys reach a $5\sigma$ galaxy depth of approximately 24.0, 23.4, and 22.5 AB magnitudes in the $g$, $r$, and $z$ bands, respectively.

\textit{The Tractor} \citep{Lang2016} performs source extraction and model-based photometry using the point-spread function and galaxy-profile fitting to deblend overlapping sources. This photometric data set, previously noted as the basis for DESI's target selection, also serves as a reference in our work.

\section{Spectroscopic Selection} \label{sec:spec}

\subsection{Pre-filtering} \label{subsec:pre_filter}
We begin with the full DESI DR1 merged redshift catalog\footnote{\url{https://data.desi.lbl.gov/public/dr1/spectro/redux/iron/zcatalog/v1/zall-pix-iron.fits}} containing approximately 28 million spectra. From this data set, we apply initial quality filters to ensure reliability. First, we remove all spectra with `ZWARN' flags not equal to zero, indicating potential issues with the pipeline fitting. Next, we exclude spectra where the target type `TGT' is classified as a star. Finally, for each object with multiple coadded spectra, we retain only the spectrum with the longest effective exposure time. These pre-filtering steps are essential for removing unreliable measurements and focusing on extragalactic objects that are potential lens candidates, and they reduce the number of useful spectra to approximately 15.8 million.

\subsection{Grouping} \label{subsec:group}
After filtering, we apply the Friends-of-Friends (FoF) algorithm using the \textit{spherimatch}\footnote{\url{https://github.com/technic960183/spherimatch}} package \citep{Hsu2025}, developed by our team, to identify spatially clustered spectra. We set a linking length of $3\farcs0$, which groups spectra that potentially belong to a strong lensing system. This approach efficiently identifies candidate lensing systems, where multiple spectra appear in close proximity on the sky but at different redshifts.

To further refine our sample, we apply a redshift-based filter, excluding groups where the ratio of the maximum to minimum redshift is less than 1.3. This criterion is designed to identify systems with significant redshift differences, characteristic of gravitational lensing phenomena. After applying these filters, we retain 13,218 groups containing a total of 26,621 spectra. These groups are distributed as 13,044 pairs (2 spectra), 165 triplets (3 spectra), 7 quartets (4 spectra), and 2 quintets (5 spectra).

\subsection{Visual Inspection for Spectral Validation} \label{subsec:vi_spec}
To ensure the reliability of our candidate selection, we conduct a comprehensive visual inspection of all 26,621 spectra, with two primary goals: validating the pipeline-fitted redshifts and correcting cases where the pipeline redshifts are incorrect. To achieve this, we assign a quality flag to each spectrum based on the clarity and number of spectral features, retaining only those with high or moderate quality.

For spectra where we identify clear discrepancies between the spectral features and the pipeline-fitted redshift, we manually assign a corrected redshift. These spectra are then re-evaluated with the corrected redshift and graded using the same criteria as other spectra.

In parallel, we record basic spectral properties, including the presence of emission or absorption lines, and the spectral type (galaxy, quasar, or star) for reference in subsequent analysis. We then exclude the spectra belonging to stars that have been misclassified as extragalactic sources.

We use the following three-level quality grading system during this inspection:
\begin{itemize}
    \item \textbf{High}: Redshift is confidently determined with at least two strong spectral features.
    \item \textbf{Moderate}: Redshift is probable based on either one strong doublet feature or multiple weak features together with the continuum (the 4000 \AA\ break).
    \item \textbf{Reject}: Redshift is unreliable, with apparent features likely arising from artifacts or imperfect sky subtraction, or with no clear features.
\end{itemize}
In this grading system, ``strong'' features refer to those visually about 2 times stronger than local fluctuations, while ``weak'' features are between 0.5 and 2 times as strong.

\begin{figure*}
    \centering
    \includegraphics[width=\linewidth]{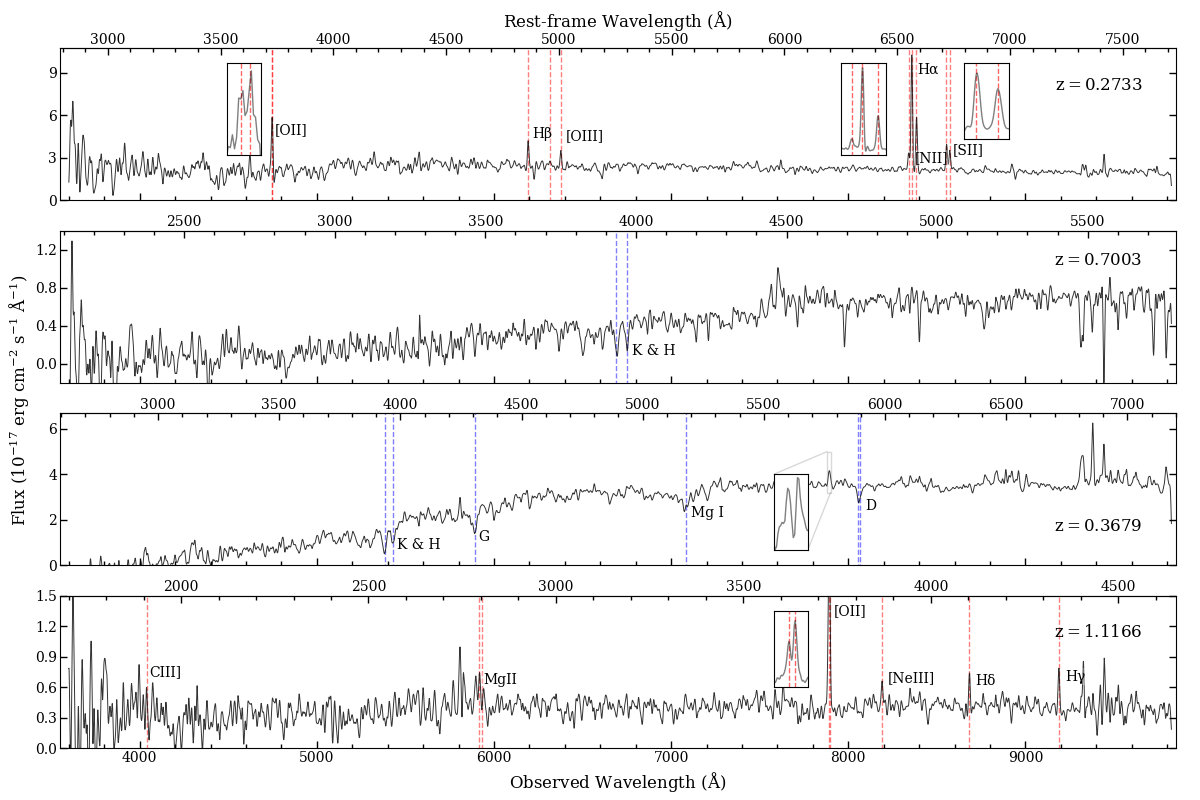}
    \caption{Representative spectra for the redshift quality grading system. Top panel: High redshift quality. The emission lines secure the redshift. From left to right, the zoomed-in boxes show the details of [\ion{O}{2}]$\lambda\lambda3726,3729$, H$\alpha$ and [\ion{N}{2}]$\lambda\lambda6550,6585$, and [\ion{S}{2}]$\lambda\lambda6718,6733$, respectively. Second panel: Moderate redshift quality. The strength of Ca H\&K absorption lines is slightly higher than the noise, and the 4000 \AA\ break is detected. Third and fourth panels: A pair of spectra showing a foreground galaxy (third panel, $z = 0.3679$) and a background galaxy (fourth panel, $z = 1.1166$). The foreground spectrum is labeled with the ``partner's Z'' flag because of the [\ion{O}{2}]$\lambda\lambda3726,3729$ emission lines from the background spectrum, visible in both panels (see the zoomed-in boxes). The background spectrum does not have the flag. Both of these spectra are graded as high quality.}
    \label{fig:spectra_examples}
\end{figure*}

Additionally, we use a separate ``partner's Z'' flag for spectra in which spectral features of another group member appear, potentially indicating a multiply imaged source. During visual inspection, we can overlay spectral line markers at an alternative redshift and compare directly with other group members. If the features match those of a partner rather than the target itself, we apply the ``partner's Z'' flag.
See Fig.~\ref{fig:spectra_examples} for examples of accepted spectra, including a case with such ``partner's Z'' features across group members.
After filtering through this visual inspection process, we obtain a refined sample of 11,837 groups comprising 23,811 spectra.

Furthermore, for groups that contain more than two spectra, we find that only 13 triplets have three well-separated redshifts ($\Delta z\geq0.02$), and the rest have only two. For the triplets that have three well-separated redshifts, we break them into three pairs of spectra. For the remaining groups, most consist of multiple fibers pointing at different parts of an object and a single fiber pointing at a second object. We select a representative pair of spectra for each group based on fiber positions or spectral quality. Pairs that do not satisfy a separation of $<3\farcs0$ and a redshift ratio of $>1.3$ are removed. With this method, we convert all groups into 11,848 pairs of spectra (hereafter ``systems'') for the second round of visual inspection in the next section.

\section{Candidate Inspection and Results} \label{sec:img}
This section presents the visual inspection of potential lensing systems using a combination of spectroscopic and imaging information. With groups of spectra having reliable redshifts established in the previous section, we assess whether their configurations exhibit features commonly associated with strong gravitational lensing.

In Section~\ref{subsec:factors}, we outline the key observational features considered during inspection. Section~\ref{subsec:grading} introduces the grading scheme used to classify candidates based on the strength and consistency of these features. The outcome of the inspection is presented in Section~\ref{subsec:result}. In Section~\ref{subsec:dimple}, we describe a new class of systems, referred to as \textit{dimple systems}, that exhibit less conventional but likely lensing-related signatures.

\subsection{Visual Inspection to Determine Lens Candidacy} \label{subsec:factors}
The visual inspection to determine lens candidacy is guided by several observational factors drawn from both imaging and spectroscopy, including:

\begin{itemize}
  \item \textbf{Arc-like or elongated features} that curve toward the central object.
  \item \textbf{Multiple or counter-image configurations}, as well as rings, typically exhibiting similar colors.
  \item \textbf{Deviation from PSF-like morphology} in quasar sources, which may suggest unresolved multiple images. This is analogous to ``elongated features'' in the first bullet.
  \item \textbf{Spectral features of sources} appearing in the spectrum of the lens (referred to as the ``partner's Z'' in Section~\ref{subsec:vi_spec}). This is considered potential evidence for a counter-image.
  \item \textbf{An estimated Einstein radius}, derived from the velocity dispersion of the lens assuming an isothermal mass profile \citep{Narayan1996}. This provides an angular scale for comparison with the positions of lensed features (see Fig.~\ref{fig:estimatedER_example} for an example). The angular Einstein radius $\theta_E$ is calculated using:
  \begin{equation}
  \theta_E = 4\pi \left( \frac{\sigma_v}{c} \right)^2 \frac{D_{\mathrm{ds}}}{D_{\mathrm{s}}},
  \end{equation}
  where $\sigma_v$ is the velocity dispersion of the lens, $c$ is the speed of light, $D_{\mathrm{s}}$ is the angular diameter distance to the source, and $D_{\mathrm{ds}}$ is the angular diameter distance between the lens and source. All distances are computed assuming a flat $\Lambda$CDM cosmology with $H_0 = 70\ \mathrm{km\ s^{-1}\ Mpc^{-1}}$ and $\Omega_m = 0.3$, and $\sigma_v$ is obtained from the \texttt{FastSpecFit} Spectral Synthesis and Emission-Line Catalog\footnote{\url{https://data.desi.lbl.gov/public/dr1/vac/dr1/fastspecfit/iron/v2.1/catalogs/fastspec-iron.fits} and its \href{https://data.desi.lbl.gov/doc/releases/dr1/vac/fastspecfit/}{website}}.

  \begin{figure}
      \centering
      \includegraphics[width=\linewidth]{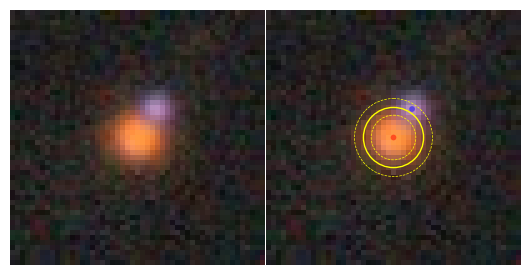}
      \caption{Images used for visual inspection. The same image is shown twice; the right panel includes annotations, while the left panel does not, to avoid obscuring faint features. In the right panel, the yellow solid circle marks the estimated Einstein radius, and the yellow dashed circles show its 1-sigma uncertainty. The red and blue dots indicate fiber positions for the foreground and background objects, respectively. The system shown here is DESI-043.3037+03.0729 (see \S~\ref{subsub:newA}).}
      \label{fig:estimatedER_example}
  \end{figure}

\end{itemize}

These factors are considered together during inspection. The combination of spectral evidence and morphological features from imaging often provides stronger support for lensing candidacy than from either the spectral or imaging data alone.
When considering only one of these two aspects, the evidence may be ambiguous; for example, a pair of foreground and background objects in proximity on the sky with different redshifts may merely overlap (without the strong lensing effect present); the spectral flag ``partner's Z'' may result from the light of the background object ``leaking'' into the fiber centered on the foreground object due to finite seeing; or an elongated feature in imaging may simply be a spiral arm, clarified by the spectra.
Our grading reflects a holistic judgment that also considers the plausibility of alternative explanations. To minimize bias, we do not check whether a system is known from previous studies before the completion of the visual inspection.

\subsection{Candidate Grading} \label{subsec:grading}
We evaluate each system to determine whether it meets the criteria for one of three grade levels: A, B, or C. Systems that receive any of these grades are selected as lens candidates. The grading reflects the strength and consistency of the observed lensing features, as detailed below.

\paragraph{Grade A:}\begin{itemize}
\item \textbf{With an estimated Einstein radius (from velocity dispersion) available:} Candidates display multiple images, an arc-counterarc configuration, or a significant elongation with an extent that is consistent with the estimated Einstein radius.
\end{itemize}

\paragraph{Grade B:}\begin{itemize}
\item \textbf{With an estimated Einstein radius:} Candidates exhibit elongation or faint multiple/counter images, with an extent approximately consistent with the estimated Einstein radius; alternatively, they show multiple/counter images with the ratio of the separation of the fibers to the expected Einstein radius being $\gtrsim1.5$.
\item \textbf{Without an estimated Einstein radius:} Candidates show multiple images or an arc-counterarc pair.
\end{itemize}

\paragraph{Grade C:}\begin{itemize}
\item \textbf{With an estimated Einstein radius:} Candidates exhibit possible elongation with an extent that is approximately consistent with the expected Einstein radius; alternatively, they show elongation, faint multiple/counter images with the ratio of the separation of the fibers to the expected Einstein radius being $\gtrsim1.5$.
\item \textbf{Without an estimated Einstein radius:} Candidates display possible elongation or multiple images, paired with a luminous (typically red) foreground galaxy.
\end{itemize}

\subsection{Conventional Strong Lens Candidates} \label{subsec:result}
From the visual inspection of 11,848 systems, we identify 2046 conventional lens candidates, of which 1906 are newly discovered and 140 are previously known systems independently recovered by our method (see Section~\ref{subsec:bias} for discussion). These known systems are originally reported by \citet{Bolton2008, Lacki2009, Brownstein2012, Sonnenfeld2013, Gavazzi2014, More2016a, Sonnenfeld2018, Lemon2018, Wong2018, Manjon-Garcia2019, Lemon2019, Petrillo2019, Jacobs2019, Gonzalez-Nuevo2019, Sonnenfeld2019, Chan2020b, Huang2020, Jaelani2020, Sonnenfeld2020, Canameras2020, Huang2021, Talbot2021, Canameras2021, Li2021, Stein2022, Shu2022, Wong2022, Savary2022, Storfer2024}. The distribution of grades is summarized in Table~\ref{tab:trad_sum}.

\begin{deluxetable}{lrrrr}
\tablewidth{\linewidth}
\tablecaption{Grading Results for Conventional Lens Candidates \label{tab:trad_sum}}
\tablehead{
\colhead{Grade} & \colhead{Grade A} & \colhead{Grade B} & \colhead{Grade C} & \colhead{\textbf{Total}}
}
\startdata
Newly Discovered  & 20  & 188  & 1698 & 1906 \\
Previously Known  & 41  &  45  &   54 &  140 \\
\hline
\textbf{Total} & \textbf{61} & \textbf{233} & \textbf{1752} & \textbf{2046} \\
\enddata
\end{deluxetable}

\begin{figure*}
    \centering
    \includegraphics[width=\linewidth]{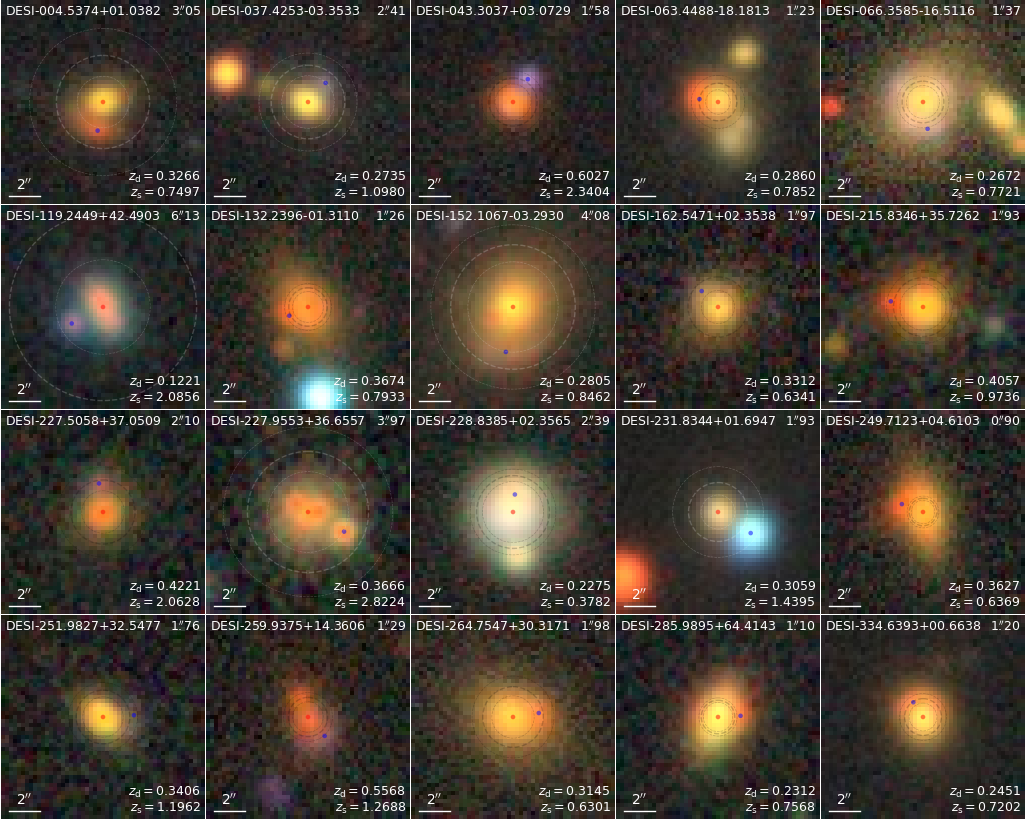}
    \caption{All 20 new Grade A candidates discovered in this paper. The naming convention is R.A. and decl. in decimal format. The images have a width of 51 pixels $\approx 13\farcs4$. North is up, and east to the left. The gray solid circle marks the estimated Einstein radius (denoted in the upper right of each image), and the gray dashed circles indicate its 1-sigma uncertainty. The red and blue dots indicate fiber positions for the foreground and background objects, respectively. $z_\mathrm{d}$ and $z_\mathrm{s}$ are the spectroscopic redshifts of the lens and the source, respectively. See \S~\ref{subsub:newA} for the description of each candidate.}
    \label{fig:newA}
\end{figure*}

All newly discovered Grade A lens candidates are presented in Figure~\ref{fig:newA}, with descriptions in Section~\ref{subsub:newA} and detailed properties listed in Table~\ref{tab:newA}. All other candidates are summarized in Appendix~\ref{app:full_cat}, which includes a machine-readable catalog with a descriptive table. Image cutouts of all new Grade B candidates are shown in Appendix~\ref{app:img_newB}.

\begin{table*}
\centering
\caption{Properties of New Grade A Lens Candidates}
\label{tab:newA}
\begin{tabular}{ccc|ccccc|ccccc}
\toprule
Name & $\theta_\mathrm{DS}$ & $\theta_\mathrm{E}$ & \multicolumn{5}{c|}{Lens} & \multicolumn{5}{c}{Source} \\
  &   &   & $z_{\mathrm{d}}$ & $\sigma_v$ & $g$ & $r$ & $z$ & $z_{\mathrm{s}}$ & Type & $g$ & $r$ & $z$ \\
\hline
\texttt{DESI-004.5374+01.0382} & $1.91$ & ${3.05}^{+1.76}_{-1.36}$ & 0.327 & $455\pm116$ & $20.96$ & $19.36$ & $18.47$ & 0.750 & GAL & $23.63$ & $22.39$ & $20.73$ \\
\texttt{DESI-037.4253-03.3533} & $1.70$ & ${2.41}^{+0.80}_{-0.68}$ & 0.274 & $348\pm53$ & $20.64$ & $19.32$ & $18.56$ & 1.098 & GAL & $22.80$ & $22.95$ & $22.43$ \\
\texttt{DESI-043.3037+03.0729} & $1.78$ & ${1.58}^{+0.47}_{-0.41}$ & 0.603 & $300\pm41$ & $21.89$ & $20.41$ & $19.08$ & 2.340 & QSO & $21.67$ & $21.63$ & $20.99$ \\
\texttt{DESI-063.4488-18.1813} & $1.23$ & ${1.23}^{+0.52}_{-0.43}$ & 0.286 & $270\pm52$ & $20.12$ & $18.67$ & $17.71$ & 0.785 & GAL & N/A & $23.68$ & $20.72$ \\
\texttt{DESI-066.3585-16.5116} & $1.79$ & ${1.37}^{+0.32}_{-0.29}$ & 0.267 & $280\pm31$ & $19.07$ & $17.88$ & $17.10$ & 0.772 & GAL & $22.22$ & $22.53$ & $21.99$ \\
\texttt{DESI-119.2449+42.4903} & $2.31$ & ${6.13}^{+4.07}_{-3.04}$ & 0.122 & $485\pm140$ & $20.05$ & $19.30$ & $18.63$ & 2.086 & QSO & $21.87$ & $21.73$ & $21.37$ \\
\texttt{DESI-132.2396-01.3110} & $1.35$ & ${1.26}^{+0.27}_{-0.25}$ & 0.367 & $301\pm31$ & $20.85$ & $19.09$ & $17.85$ & 0.793 & GAL & $26.25$ & $23.69$ & $21.33$ \\
\texttt{DESI-152.1067-03.2930} & $2.98$ & ${4.08}^{+1.30}_{-1.12}$ & 0.281 & $480\pm71$ & $19.48$ & $17.89$ & $16.94$ & 0.846 & GAL & $22.32$ & $21.86$ & $21.09$ \\
\texttt{DESI-162.5471+02.3538} & $1.49$ & ${1.97}^{+1.14}_{-0.88}$ & 0.331 & $396\pm102$ & $20.85$ & $19.32$ & $18.44$ & 0.634 & GAL & $23.54$ & $23.16$ & $22.59$ \\
\texttt{DESI-215.8346+35.7262} & $2.14$ & ${1.93}^{+0.47}_{-0.42}$ & 0.406 & $361\pm42$ & $20.76$ & $18.87$ & $17.84$ & 0.974 & GAL & N/A & $24.15$ & $21.15$ \\
\texttt{DESI-227.5058+37.0509} & $1.89$ & ${2.10}^{+0.85}_{-0.71}$ & 0.422 & $325\pm60$ & $21.16$ & $19.50$ & $18.55$ & 2.063 & QSO & $23.00$ & $22.90$ & $22.10$ \\
\texttt{DESI-227.9553+36.6557} & $2.69$ & ${3.97}^{+1.59}_{-1.32}$ & 0.367 & $423\pm78$ & $20.45$ & $19.10$ & $18.10$ & 2.822 & QSO & $21.21$ & $20.06$ & $19.17$ \\
\texttt{DESI-228.8385+02.3565} & $1.15$ & ${2.39}^{+0.42}_{-0.39}$ & 0.228 & $470\pm40$ & $19.06$ & $18.25$ & $17.70$ & 0.378 & QSO & $19.28$ & $18.31$ & $17.42$ \\
\texttt{DESI-231.8344+01.6947} & $2.55$ & ${1.93}^{+1.04}_{-0.82}$ & 0.306 & $306\pm74$ & $20.95$ & $19.91$ & $19.30$ & 1.440 & QSO & $19.15$ & $18.66$ & $18.81$ \\
\texttt{DESI-249.7123+04.6103} & $1.48$ & ${0.90}^{+0.14}_{-0.13}$ & 0.363 & $284\pm21$ & $20.67$ & $18.94$ & $17.86$ & 0.637 & GAL & N/A & $24.60$ & $20.95$ \\
\texttt{DESI-251.9827+32.5477} & $2.03$ & ${1.76}^{+0.50}_{-0.43}$ & 0.341 & $308\pm41$ & $20.97$ & $19.39$ & $18.51$ & 1.196 & GAL & $23.30$ & $23.26$ & $22.61$ \\
\texttt{DESI-259.9375+14.3606} & $1.65$ & ${1.29}^{+0.32}_{-0.28}$ & 0.557 & $308\pm36$ & $22.16$ & $20.46$ & $19.12$ & 1.269 & GAL & $22.52$ & $22.44$ & $21.80$ \\
\texttt{DESI-264.7547+30.3171} & $1.71$ & ${1.98}^{+0.70}_{-0.59}$ & 0.315 & $388\pm63$ & $20.02$ & $18.40$ & $17.45$ & 0.630 & GAL & $25.02$ & $23.27$ & $21.09$ \\
\texttt{DESI-285.9895+64.4143} & $1.48$ & ${1.10}^{+0.15}_{-0.14}$ & 0.231 & $242\pm16$ & $19.90$ & $18.45$ & $17.55$ & 0.757 & GAL & $25.46$ & $23.32$ & $20.88$ \\
\texttt{DESI-334.6393+00.6638} & $1.15$ & ${1.20}^{+0.33}_{-0.29}$ & 0.245 & $260\pm34$ & $20.41$ & $18.92$ & $17.99$ & 0.720 & GAL & $25.18$ & $23.75$ & $20.88$ \\
\hline
\end{tabular}
\tablecomments{The naming convention is R.A. and decl. in decimal format. $\theta_{\mathrm{DS}}$ is the angular separation between the fiber pair on the sky (between the putative lens and the source image), and $\theta_{\mathrm{E}}$ is the estimated Einstein radius, both in arcseconds. For the lens, $z_\mathrm{d}$ is the spectroscopic redshift, $\sigma_v$ is the velocity dispersion in km\,s$^{-1}$, and $g$, $r$, $z$ are the $g$, $r$, $z$-band magnitudes, respectively. For the source, $z_\mathrm{s}$ is the spectroscopic redshift, type is the spectral type, and $g$, $r$, $z$ are also the magnitudes.}
\end{table*}

\subsubsection{Description of New Grade A Candidates} \label{subsub:newA}
\begin{figure*}
    \centering
    \includegraphics[width=\linewidth]{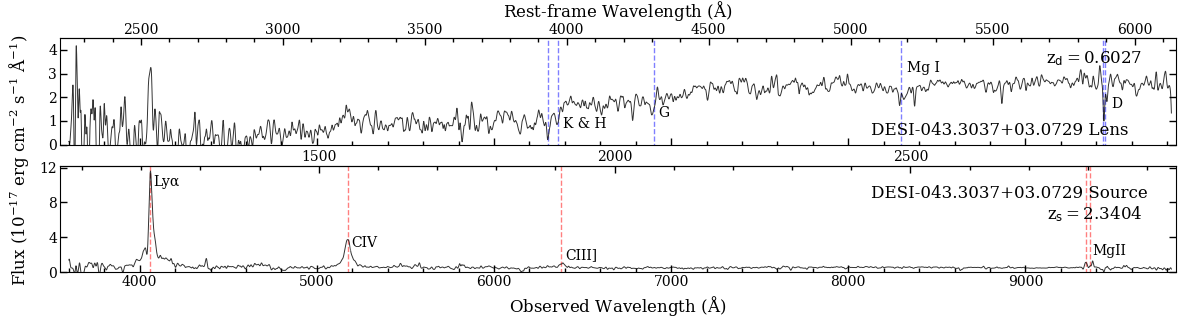}
    \caption{Spectra of DESI-043.3037+03.0729 with spectroscopic redshift $z_\mathrm{d} = 0.6027$ (lens, top) and $z_\mathrm{s} = 2.3404$ (source, bottom). The emission lines Ly$\alpha$ and \ion{C}{4}$\lambda1549$ from the source appear in the lens spectrum.}
    \label{fig:newA_spec1}
\end{figure*}

\begin{figure*}
    \centering
    \includegraphics[width=\linewidth]{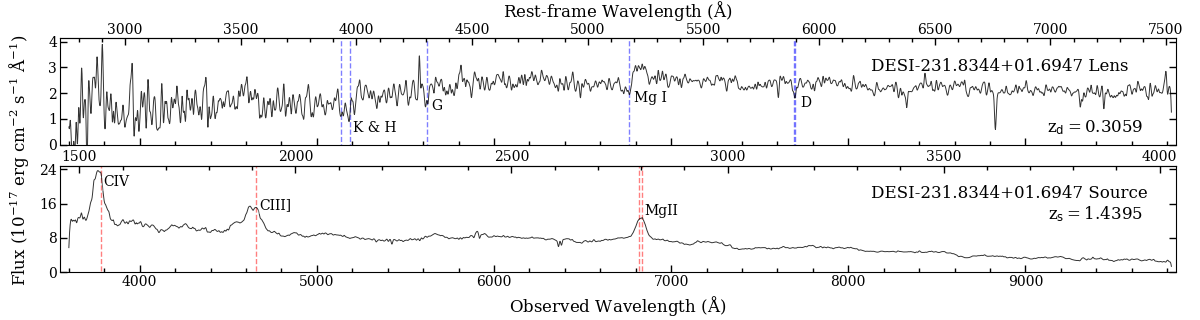}
    \caption{Spectra of DESI-231.8344+01.6947 with spectroscopic redshift $z_\mathrm{d} = 0.3059$ (lens, top) and $z_\mathrm{s} = 1.4395$ (source, bottom). Broad emission lines [\ion{Mg}{2}]$\lambda2799$ and \ion{C}{4}$\lambda1549$ from the source appear in the lens spectrum.}
    \label{fig:newA_spec2}
\end{figure*}

Here we provide descriptions of all newly discovered Grade A candidates. The candidates are listed in the same order as in Figure~\ref{fig:newA} and Table~\ref{tab:newA}. Pairs of spectra for two selected candidates are provided in Figures~\ref{fig:newA_spec1} and \ref{fig:newA_spec2}.

\paragraph{DESI-004.5374+01.0382} A large arc in the SE and a counter-image in the W.
\paragraph{DESI-037.4253-03.3533} An image in the NW and a counter-image in the S. The [\ion{O}{2}]$\lambda\lambda3726,3729$ doublet from the source appears in the lens spectrum.
\paragraph{DESI-043.3037+03.0729} A doubly lensed quasar system. Both Ly$\alpha$ and \ion{C}{4}$\lambda1549$ emission lines from the source appear in the lens spectrum---these appear as broad features, though not as broad as for a typical QSO (see Figure~\ref{fig:newA_spec1}).
\paragraph{DESI-063.4488-18.1813} A large arc in the E and a counter-arc in the W.
\paragraph{DESI-066.3585-16.5116} A quadruply lensed system. The [\ion{O}{2}]$\lambda\lambda3726,3729$ and [\ion{O}{3}]$\lambda\lambda4959,5007$ doublets, as well as the H$\beta$ emission line from the source, appear in the lens spectrum.
\paragraph{DESI-119.2449+42.4903} A slightly elongated quasar image in the SE and a counter-image in the SW.
\paragraph{DESI-132.2396-01.3110} An arc in the E and a counter-arc in the W. Two possible images in the NW and the SW make this system a possible quad.
\paragraph{DESI-152.1067-03.2930} A large arc in the S.
\paragraph{DESI-162.5471+02.3538} A large arc in the NE and a possible counter-arc in the SW.
\paragraph{DESI-215.8346+35.7262} An image in the E and another in the N; possibly a quad.
\paragraph{DESI-227.5058+37.0509} A doubly lensed quasar system. An image in the N and a counter-image in the SE.
\paragraph{DESI-227.9553+36.6557} A possible quadruply lensed quasar system with three clear images and a possible fourth image in the S. The \ion{C}{3}]$\lambda1909$ broad emission line from the source appears faintly in the lens spectrum.
\paragraph{DESI-228.8385+02.3565} A doubly lensed quasar system. The spectrum of the source comes from the image in the N. The image in the S does not have a DESI spectrum at present. Features from both spectra appear in each other, likely due to the small separation between the fibers.
\paragraph{DESI-231.8344+01.6947} A lensed quasar system. There appears to be a very faint arc around the lens, starting from the bright image in the SW and extending toward the N. In addition, the [\ion{Mg}{2}]$\lambda2799$ and \ion{C}{4}$\lambda1549$ broad emission lines from the source appear in the lens spectrum (see Figure~\ref{fig:newA_spec2}).
\paragraph{DESI-249.7123+04.6103} A large arc in the NE and a possible counter-arc in the SW.
\paragraph{DESI-251.9827+32.5477} A large but faint arc in the W and another arc of similar size in the E.
\paragraph{DESI-259.9375+14.3606} A large arc in the SW and a possible counter-arc in the NE. The [\ion{O}{2}]$\lambda\lambda3726,3729$ emission lines from the source appear in the lens spectrum.
\paragraph{DESI-264.7547+30.3171} An arc in the W and a possible counter-arc in the E.
\paragraph{DESI-285.9895+64.4143} An arc in the W and a counter-arc in the E.
\paragraph{DESI-334.6393+00.6638} Two arcs in the NE and the NW, with two other possible counter-arcs on the opposite side. Features from both spectra appear in each other, likely due to the small separation between the fibers.

\subsection{Dimple Systems} \label{subsec:dimple}
During visual inspection, we also identified a class of lens candidates that we refer to as ``dimple systems.'' In these cases, a small, low-mass foreground galaxy, acting as a lens, appears superimposed on the extended light profile of a large background galaxy. While the lens typically has a small Einstein radius and does not produce obvious arcs or counter-images, it may induce a localized indentation, visually resembling a dimple, in the surface brightness profile of the background source (see the simulations of such systems in \citet{Silver2025}).

\begin{figure*}
    \centering
    \includegraphics[width=\linewidth]{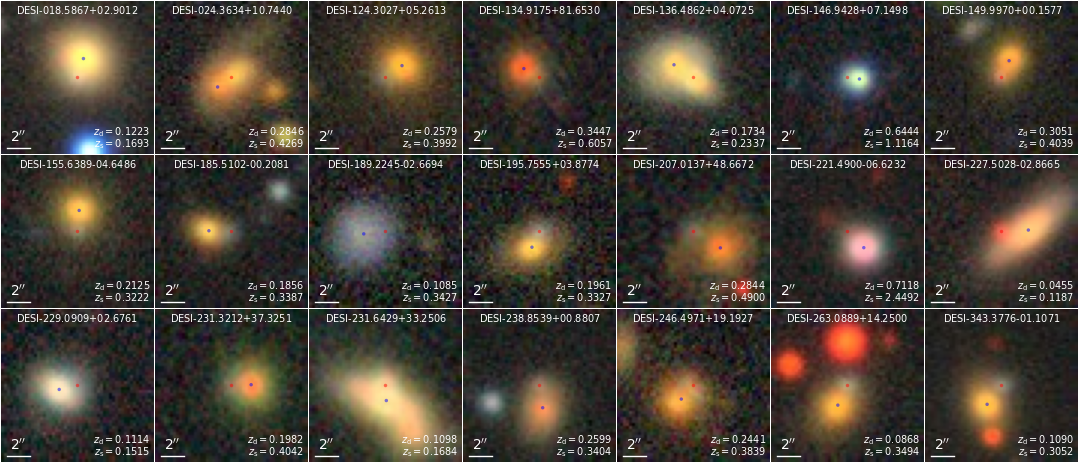}
    \caption{Representative ``dimple'' lens candidates discovered in this paper. The image format and annotations follow the same convention as in Figure~\ref{fig:newA}. Velocity dispersion is not available for most of the dimple candidates; thus the estimated Einstein radius is not provided.}
    \label{fig:dimple_exp}
\end{figure*}

We have identified 318 dimple candidates (some overlapping with the conventional lens candidates); examples are shown in Figure~\ref{fig:dimple_exp}.

Follow-up observations of these candidate dimple systems are needed to confirm their lensing nature. For example, high-resolution imaging can reveal the indentation in the light distribution of the background galaxy. Multi-band imaging may offer improved contrast between the lens and the background source.
In addition, a data cube from an integral field unit can reveal the relative strength of the emission-line fluxes of the background galaxy with respect to the location of the foreground low-mass galaxy, either from space or from a ground-based observatory with adaptive optics.

\section{Discussion} \label{sec:discuss}

\subsection{The New Methodology and the Prediction}
Our method identified a total of 2046 conventional strong lens candidates (Grades A--C), including 1906 newly discovered systems and 140 previously known lenses recovered. These findings demonstrate that our method is effective for discovering new lens candidates, particularly those that may be missed by ML-based searches in imaging data or by single-fiber searches in spectroscopic data, since our method integrates both spectral (across two or more fibers) and imaging information.

\begin{figure*}[h]
    \centering
    \includegraphics[width=\linewidth]{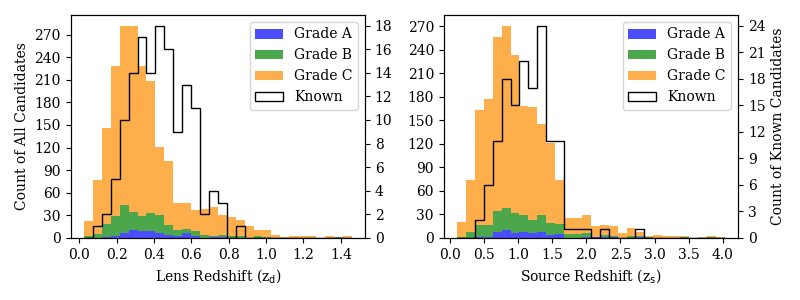}
    \caption{Redshift distributions of conventional lens candidates by grade. Left panel: Distribution of lens redshifts. Right panel: Distribution of source redshifts. In both panels, stacked colored histograms represent all conventional candidates identified in this work, grouped by visual inspection grade: Grade A (blue), Grade B (green), and Grade C (orange), corresponding to the left y-axis. The black solid lines represent the redshift distributions of previously known strong lensing systems, corresponding to the right y-axis.}
    \label{fig:zdist_convention}
\end{figure*}

The redshift distributions of these candidates are shown in Figure~\ref{fig:zdist_convention}. For the lenses, the distribution peaks at $z_\mathrm{d}\sim0.3$, which is lower than that of the known candidates (at $z_\mathrm{d}\sim0.4$), but has a slightly broader range. Grade A systems are the most consistent with the known sample, with Grade B and Grade C candidates showing progressively greater differences. The distribution of the sources also shows a similar trend.
This may indicate that our Grade B and C candidates represent lensing systems that occupy a broader parameter space than the known lenses.

\begin{figure*}
    \centering
    \includegraphics[width=\linewidth]{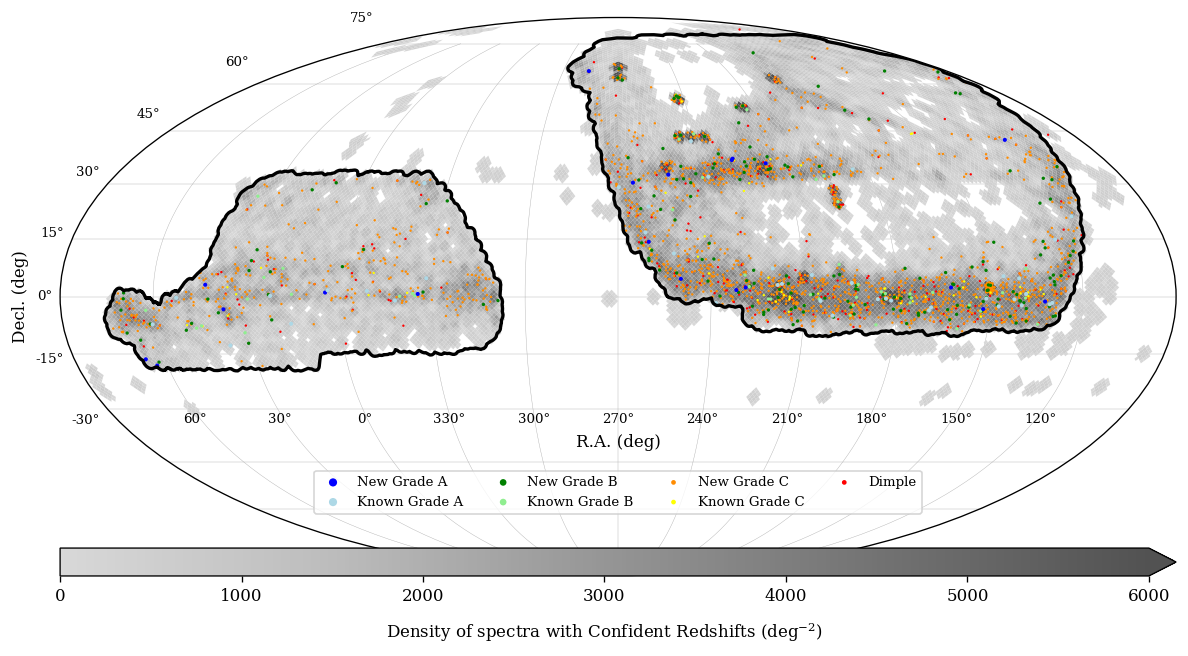}
    \caption{All candidates identified in this work overlaid on the DESI DR1 footprint. Each point represents an individual candidate with the following color scheme: newly discovered Grade A (blue), B (green), and C (dark orange); previously known Grade A (light blue), B (light green), and C (yellow); and dimple candidates (red). The grayscale background shows the density of spectra with confident redshifts (\texttt{ZWARN=0} and \texttt{OBJTYPE=TGT}). The black solid line represents the footprint of the DARK program of the DESI main survey. This map is drawn with the Mollweide projection in equatorial coordinates.}
    \label{fig:map}
\end{figure*}

Figure~\ref{fig:map} shows the spatial distribution of all visually identified candidates overlaid on the DESI DR1 spectroscopic footprint. The strong correlation between candidate density and the underlying spectroscopic coverage highlights the dependence of our method on fiber density. In particular, the small clumps of candidates in high-density regions are due to the ``One-Percent Survey'' conducted during the DESI Survey Validation phase, which followed a ``deep-first'' strategy and created localized regions of high fiber coverage. This reflects the clear advantage of increased spectroscopic density for this approach to strong-lens searches.

By performing a linear fit between the density of spectra with reliable redshift measurements in DESI DR1 $\rho_\textrm{spec}$ and the density of candidates $\rho_\textrm{can}$ on a HEALPix map with $N_\textrm{side}=16$, we obtain the following relation:
$$\rho_\textrm{can}=0.000154\times\rho_\textrm{spec}-0.0703,\,r^2=0.641$$.
This corresponds to an average yield of one strong-lens candidate per $\sim6500$ redshift measurements within a unit solid angle. (``Injected'' candidates are not included in this estimate; see \S~\ref{subsec:bias}.)

As DESI progresses toward its goal of obtaining redshifts for 40 million galaxies and quasars across $14{,}000\deg^2$, and given that DR1 already contains about one-third of the redshift measurements, we estimate that our method could identify at least $\sim4000$ additional lens candidates. This would bring the total number of new candidates discovered with our method to approximately 6000 by the end of the DESI 5-year main survey.
This demonstrates that our method is competitive with \emph{and} complementary to the number of lenses found or projected to be found by using imaging data \citep{Huang2020, Huang2021, Storfer2024, Inchausti2025} or the single-fiber search technique (J.~Karp et al. in prep.) on the same DESI footprint.
Not counting the single-fiber search that is expected to be published soon, there are now $\sim10,000$ lens candidates from DESI imaging and spectroscopic data.
Combining all search methods, we expect the number of strong lenses (with redshifts for the lenses and sources for most systems) from the DESI footprint will be of the same order of magnitude as that projected to be found in the Rubin Observatory Legacy Survey of Space and Time (LSST). Together with UNIONS\footnote{The MzLS part of the DESI Legacy Surveys covers approximately the same footprint. However, the image quality is inferior to UNIONS.} \citep{Storfer2025}, this sample will provide a highly complementary set of lenses, with LSST uncovering systems in the southern sky and 4MOST enabling their follow-up.

\subsection{Dimple Candidates with Dwarf Galaxies}
\begin{figure*}
    \centering
    \includegraphics[width=\linewidth]{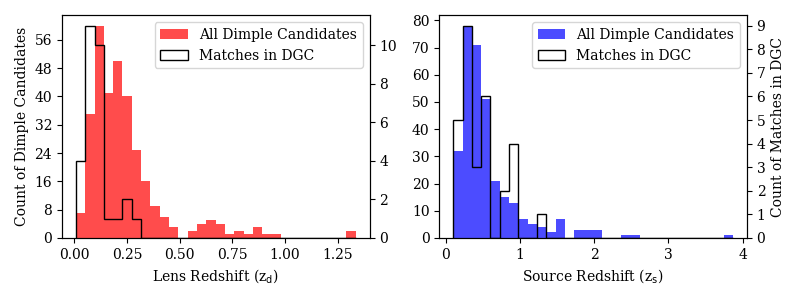}
    \caption{Redshift distributions of dimple candidates. Left panel: Distribution of lens redshifts. Right panel: Distribution of source redshifts. In both panels, the filled histograms (red for lenses, blue for sources) show the redshift distributions of all dimple candidates identified in this work, corresponding to the left y-axis. The black solid lines represent the redshift distributions of systems matched with the DESI Extragalactic Dwarf Galaxy Catalog (DGC), corresponding to the right y-axis.}
    \label{fig:zdist_dimple}
\end{figure*}

In addition to the conventional candidates, we identify 318 systems characterized by lensing configurations with small Einstein radii. We interpret these systems as potential candidates for strong lensing by dwarf galaxies. To explore this connection, we cross-match our dimple candidates with the DESI Extragalactic Dwarf Galaxy Catalog\footnote{\url{https://data.desi.lbl.gov/public/dr1/vac/dr1/extragalactic-dwarfs/v1.0/desi_dwarfs_y1_catalog.fits} and its \href{https://data.desi.lbl.gov/doc/releases/dr1/vac/extragalactic-dwarfs/}{website}} (DGC, Manwadkar et al. in prep.) and find 30 matches. See Appendix~\ref{app:img_dwarf} for their image cutouts.
As shown in Figure~\ref{fig:zdist_dimple}, the distribution of the putative lenses in the candidate dimple lensing systems extends to higher redshifts than the dwarf galaxies in the DESI catalog. This is not a surprise, as dwarf galaxies are difficult to identify beyond the local universe. We suspect a large number of our dimple lens candidates will turn out to be dwarf galaxies.
Therefore, our discoveries of these dimple lens candidates present a possible new avenue to find and study dwarf galaxies. First, dwarf galaxies acting as lenses will allow us to measure the total mass within the Einstein radius. The comparison with the baryonic mass will make it possible to constrain the mass of the dark matter component, and this will address the question of the nature of dark matter by constraining their mass-profile slope \citep[e.g.,][]{Zhang2024, He2025}, stellar mass-halo mass relation \citep[e.g.,][]{Read2017, Wechsler2018}, and halo mass function for $M_{\textrm{Halo}} \lesssim 10^{13}\,M_\odot$ \citep[e.g.,][]{Driver2022}. Secondly, this will allow us to identify dwarf galaxies at cosmological distances and will further allow us to examine any evolutionary trends. Lastly, by the end of the DESI program, we expect to find a total of at least $\sim1000$ dimple lens candidates. This will then constitute a statistical sample of dwarf galaxies, for which we can measure the masses reliably via the lensing effect.

\subsection{Biases from Spectral Targeting} \label{subsec:bias}
To evaluate selection biases in our method, we examine the distribution of DESI target classes that led to the initial spectroscopy of each candidate in our sample. Table~\ref{tab:dis_targeting} summarizes this for both lenses and sources across all conventional candidates, including distinctions between previously known lens candidates and newly identified ones.
Since we intentionally avoid imposing priors on the spectral type of the lens or source, this enables us to identify candidates beyond specific combinations of spectral types \citep[e.g., LRG as foreground and ELG as background from single fiber lens searches; e.g.,][]{Talbot2021}, offering a more complementary perspective on the lensing population.

\begin{table}[h]
\centering
\caption{Distribution of DESI Target Classes}
\label{tab:dis_targeting}
\begin{tabular}{l|rrr|rrr}
\toprule
Class & \multicolumn{3}{c|}{Lens} & \multicolumn{3}{c}{Sources} \\
 & New & Known & All & New & Known & All \\
\hline
BGS & 1474 & 92 & 1566 & 132 & 2 & 134 \\
LRG & 442 & 70 & 512 & 662 & 18 & 680 \\
ELG & 72 & 0 & 72 & 846 & 37 & 883 \\
QSO & 38 & 0 & 38 & 362 & 7 & 369 \\
SCND & 177 & 87 & 264 & 223 & 83 & 306 \\
- LENS & 0 & 78 & 78 & 0 & 80 & 80 \\
- others & 189 & 15 & 204 & 283 & 6 & 289 \\
\hline
\end{tabular}
\tablecomments{Each target may be selected by multiple target classes, so the total counts exceed the number of candidates. LENS refers to the \texttt{STRONG\_LENS} program. Counts from all other SCND programs are combined under ``others''.}
\end{table}

To trace the origins of each candidate, we further examine the targeting bitmasks recorded for each fiber. In addition to the main survey (e.g., BGS, LRG, ELG), we also track whether a fiber has been selected by any of the Secondary Target Programs (SCND), a set of programs that utilize remaining fibers not assigned to primary targets, thereby enabling broader scientific exploration beyond the main programs.

We find that all previously known lens candidates in our sample are associated with BGS, LRG, or SCND targets. Notably, 82 out of 140 contain at least one spectrum solely observed through the program ``Spectroscopic confirmation of strong gravitational lens candidates'' \citep[hereafter \texttt{STRONG\_LENS},][]{Huang2025}, which is specifically designed to acquire spectra of systems already identified as lens candidates from the DESI Legacy Surveys imaging data. While all spectra are treated equally in our analysis, these cases represent sample injection and therefore cannot be regarded as independent recoveries by our method.

In contrast, the newly identified candidates are mostly targeted via the main DESI survey programs, particularly BGS and LRG for the lenses, and ELG for the sources. None of these new candidates are associated with the \texttt{STRONG\_LENS} program, consistent with the expectation that they were not pre-selected as known lens candidates.

This distinction between pre-selected and independently discovered candidates underscores the importance of separating sample injection from genuine recovery. The targeting breakdown demonstrates that our pipeline is capable of recovering known lenses and, crucially, identifying new systems from the main DESI survey without relying on prior lensing information. This motivates a deeper analysis of recovery efficiency using only the subset of previously known candidates unaffected by targeted injection, which we present in the following subsection.

\subsection{Recovery of Previously Known Candidates} \label{subsec:recovery}
To further quantify the performance of our methodology, we evaluate how previously known lens candidates (excluding those solely observed through the \texttt{STRONG\_LENS} program as mentioned above) progress through each stage of our pipeline. Out of 3619 such systems with at least one DESI fiber, only 154 have multiple fibers within 3 arcseconds, a necessary condition for forming spectral groups in this work. Among these, 90 systems have reliable redshifts in two or more spectra after visual inspection, and 78 of them satisfy the redshift ratio threshold. Ultimately, 56 pass our final lens inspection. This much-reduced number (from 3619) reflects both the stringency of our selection criteria and observational limitations, particularly fiber density, which appears to be the dominant bottleneck. It is worth emphasizing that despite this limitation, DESI remains the only wide-field spectroscopic survey with sufficient coverage and fiber density to enable a pair-based lens search of this nature.

\begin{figure}
    \centering
    \includegraphics[width=\linewidth]{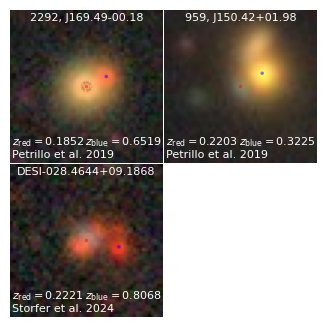}
    \caption{Rejected previously identified lens candidates during the final candidate inspection. The formatting is the same as in Figure~\ref{fig:newA}. $z_\mathrm{red}$ and $z_\mathrm{blue}$ are the spectroscopic redshifts of the foreground object (red dot) and the background object (blue dot), respectively.
    Upper left: The background image does not show elongation and its slightly asymmetric appearance may be caused by obscuration from the foreground galaxy. The estimated Einstein radius is three times smaller than the separation between the two objects.
    Upper right: The putative lens (blue dot) is in fact in the background. Note that the galaxy north of the blue-dot galaxy does not have a DESI spectrum.
    Lower left: The spectrum of the putative source (red dot) shows that it is in the foreground.
    }
    \label{fig:klc_reject}
\end{figure}

\begin{figure}
    \centering
    \includegraphics[width=0.5\linewidth]{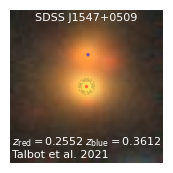}
    \caption{A previously identified lens candidate that we do not recover in this work. The formatting and annotations are the same as in Figure~\ref{fig:klc_reject}.
    \citet{Talbot2021} identified this candidate by the detection of the [\ion{O}{2}]$\lambda\lambda3726,3729$ doublet at $z=0.88$ within the $z=0.361$ galaxy (blue dot), though the source is unresolved in imaging. In our analysis, this system is selected because of the nearby foreground galaxy (red dot), at $z_\mathrm{red} = 0.2552$, forming a pair of galaxies with differing redshifts. However, it fails the final candidate inspection because the pair does not resemble a strong lensing system.
    }
    \label{fig:klc_unrecoverable}
\end{figure}

In the final stage, out of the 78 systems, 22 do not pass inspection.
14 of these are from \citet{Petrillo2019}, 4 from SuGOHI \citep{Sonnenfeld2018, Wong2018, Sonnenfeld2020}, 2 from \citet{Talbot2021}, 1 from \citet{Stein2022}, and 1 from \citet{Storfer2024}.
We find that, as shown in Figure~\ref{fig:klc_reject}, in several cases, the systems are rejected as strong lens candidates because their estimated Einstein radii are too small compared to the separation between the putative lens and source images, lowering our confidence below the threshold used in this work. In other cases, the previously presumed background source is actually a foreground galaxy.
Finally, for some systems (e.g., Figure~\ref{fig:klc_unrecoverable}), a nearby galaxy with a different redshift can lead to a spurious pair being selected, even though the actual source of the background emission is unresolved in imaging.

\section{Summary} \label{sec:summary}
In this work, we present a new method for searching for strong gravitational lenses. We identify lens candidates by matching pairs of spectra that are close in angular separation on the sky but have sufficiently large redshift differences in a spectroscopic survey. This method complements ML-based lens searches in imaging data and single-fiber lens searches in spectroscopic data. Our approach utilizes spectroscopic data from DESI DR1 and imaging from the DESI Legacy Surveys DR10. To date, DESI DR1 contains the largest publicly available redshift catalog, with 14.6 million reliable redshift measurements of galaxies and quasars. By leveraging the unprecedented fiber density of DESI, we perform a pairwise spectroscopic lens search for the first time.

Visual inspection plays a crucial role in our method. It consists of two parts: pipeline redshift validation for 26,621 spectra and image inspection for 11,848 systems.

To select the spectra for inspection, we begin with galaxies and quasars with redshifts labeled as reliable by the pipeline, then group them using our implementation \citep[spherimatch,][]{Hsu2025} of the FoF algorithm in spherical coordinates with a linking length of $3\farcs0$. We keep the groups whose maximum and minimum pipeline redshifts yield a redshift ratio greater than 1.3. We then grade each spectrum's pipeline redshift reliability as high, moderate, or reject, and proceed only with those having high or moderate grades. For spectra with clear mismatches between their features and the pipeline redshift, we manually assign corrected redshifts before grading. For groups with more than two spectra, we break them into pairs (if there are at least three well-separated redshifts in the group) or select the best pair to represent the group.

We conduct one more round of visual inspection, which involves both imaging and spectral data. On the imaging side, we consider both classical imaging features and the expected Einstein radius as grading factors. Einstein radii are estimated using the redshifts of the putative lens and source for each pair and assuming a singular isothermal mass profile. On the spectral side, we determine whether spectral features of the putative source appear in the lens spectrum, which may indicate the presence of counter-images unresolved in the imaging data.

We then assign grades A, B, or C to our candidates based on all the factors mentioned above. We present 2046 strong lens candidates. Among them, 1906 are new discoveries (including 20 new A-grade and 188 new B-grade systems) and 140 correspond to previously published lenses, of which 78 are DESI \texttt{STRONG\_LENS} secondary targets.
To our knowledge, our new candidate sample represents the largest number of strong lens candidates identified in spectroscopic observations.

In addition to conventional systems, we identify a new class of candidates, which we term dimple systems. These involve small, low-mass galaxies in front of background galaxies with extended light profiles, producing subtle surface brightness indentations rather than prominent arcs or counter-images. We identify 318 dimple systems and match 30 of them with the DESI Extragalactic Dwarf Galaxy Catalog.
This represents an alternative and complementary approach to using ML techniques to find low-mass galaxies \citep{Silver2025}.
One significant advantage of our approach is that our candidates already have redshifts for the foreground and background galaxies.
This enables the use of dwarf galaxy lenses to test predictions of the CDM model, by constraining their mass profiles, stellar mass-halo mass relation, and abundance.
However, high-resolution imaging is required for full confirmation and lens modeling to determine the enclosed mass, the slope of the mass profile, and other properties.

By overlaying the sky distribution of our candidates with the DESI spectroscopic density map, we find a strong correlation between the density of our candidates and the density of the spectroscopic data, indicating that our method depends heavily on the local density of available spectra.

Given the ongoing progress of the DESI 5-year main survey, we project that this method will discover approximately 7000 \emph{new} lens candidates in total.
This is competitive and highly complementary to the ML-based imaging and single-fiber spectroscopic lens searches on the same DESI footprint.

\begin{acknowledgments}
Y.M.H. acknowledges support from Ting-Wen Lan and Hsi-Yu Schive, who provided the opportunity to continue this research during his study at National Taiwan University.

X.H. acknowledges the University of San Francisco Faculty Development Fund.

This research used resources of the National Energy Research Scientific Computing Center (NERSC), a U.S. Department of Energy Office of Science User Facility operated under Contract No. DE-AC02-05CH11231 and the Computational HEP program in The Department of Energy's Science Office of High Energy Physics provided resources through the ``Cosmology Data Repository'' project (Grant \#KA2401022).

This material is based upon work supported by the U.S. Department of Energy (DOE), Office of Science, Office of High-Energy Physics, under Contract No. DE-AC02-05CH11231, and by the National Energy Research Scientific Computing Center, a DOE Office of Science User Facility under the same contract. Additional support for DESI was provided by the U.S. National Science Foundation (NSF), Division of Astronomical Sciences under Contract No. AST-0950945 to the NSF's National Optical-Infrared Astronomy Research Laboratory; the Science and Technology Facilities Council of the United Kingdom; the Gordon and Betty Moore Foundation; the Heising-Simons Foundation; the French Alternative Energies and Atomic Energy Commission (CEA); the National Council of Humanities, Science and Technology of Mexico (CONAHCYT); the Ministry of Science, Innovation and Universities of Spain (MICIU/AEI/10.13039/501100011033), and by the DESI Member Institutions: \url{https://www.desi.lbl.gov/collaborating-institutions}.

The DESI Legacy Imaging Surveys consist of three individual and complementary projects: the Dark Energy Camera Legacy Survey (DECaLS), the Beijing-Arizona Sky Survey (BASS), and the Mayall z-band Legacy Survey (MzLS). DECaLS, BASS and MzLS together include data obtained, respectively, at the Blanco telescope, Cerro Tololo Inter-American Observatory, NSF's NOIRLab; the Bok telescope, Steward Observatory, University of Arizona; and the Mayall telescope, Kitt Peak National Observatory, NOIRLab. NOIRLab is operated by the Association of Universities for Research in Astronomy (AURA) under a cooperative agreement with the National Science Foundation. Pipeline processing and analyses of the data were supported by NOIRLab and the Lawrence Berkeley National Laboratory. Legacy Surveys also uses data products from the Near-Earth Object Wide-field Infrared Survey Explorer (NEOWISE), a project of the Jet Propulsion Laboratory/California Institute of Technology, funded by the National Aeronautics and Space Administration. Legacy Surveys was supported by: the Director, Office of Science, Office of High Energy Physics of the U.S. Department of Energy; the National Energy Research Scientific Computing Center, a DOE Office of Science User Facility; the U.S. National Science Foundation, Division of Astronomical Sciences; the National Astronomical Observatories of China, the Chinese Academy of Sciences and the Chinese National Natural Science Foundation. LBNL is managed by the Regents of the University of California under contract to the U.S. Department of Energy. The complete acknowledgments can be found at \url{https://www.legacysurvey.org/acknowledgment/}.

Any opinions, findings, and conclusions or recommendations expressed in this material are those of the author(s) and do not necessarily reflect the views of the U. S. National Science Foundation, the U. S. Department of Energy, or any of the listed funding agencies.

The authors are honored to be permitted to conduct scientific research on I'oligam Du'ag (Kitt Peak), a mountain with particular significance to the Tohono O'odham Nation.
\end{acknowledgments}

\begin{contribution}
Y.M.H. developed the methodology, conducted the visual inspection, implemented the software tools, and prepared the manuscript.
X.H. supervised the research, provided training for the visual inspection, and reviewed and revised the manuscript.
C.S. and J.I. assisted with referencing the known candidates.
D.S. contributed to the development of concepts and the initial feasibility discussion.
J.M. developed \texttt{FastSpecFit}, which provided the velocity dispersion measurements.
All other authors, as DESI builders, contributed through their roles in the DESI Collaboration.
\end{contribution}

\facilities{Mayall(DESI), Mayall(Mosaic-3), Blanco(DECam), Bok(90Prime)}

\software{spherimatch \citep{Hsu2025},
          astropy \citep{AstropyCollaboration2013, AstropyCollaboration2018, AstropyCollaboration2022},
          desispec \citep{Guy2023},
          desitarget \citep{Myers2023},
          \texttt{FastSpecFit} \citep{Moustakas2023},
          Redrock (S.~Bailey et al., in prep.)
          }

\appendix

\section{Full Catalog of Lens Candidates} \label{app:full_cat}
This appendix presents the full list of lens candidates identified in this work as an online catalog. It includes all systems graded A, B, or C, newly discovered or previously known, as well as the dimple candidates. A machine-readable version is provided on our \href{https://sites.google.com/usfca.edu/neuralens/}{project website} and on \href{https://doi.org/10.5281/zenodo.17153224}{Zenodo}. Table~\ref{tab:catalog_description} describes the columns in the catalog.

\section{Images of New Grade B Candidates} \label{app:img_newB}
Figure~\ref{fig:newB} shows image cutouts of all newly discovered Grade B systems.

\section{Images of DESI DGC Matches} \label{app:img_dwarf}
Figure~\ref{fig:dwarf} shows image cutouts of all matches in the DGC with the dimple candidates.

\begin{deluxetable*}{lcl}
\digitalasset
\tablecaption{Description of Columns in the Full Candidate Catalog \label{tab:catalog_description}}
\tablehead{
\colhead{Column} & \colhead{Unit} & \colhead{Description}
}
\startdata
\texttt{Name} & --- & Name in the convention of R.A. and decl. in decimal format\\
\texttt{RAdeg} & deg & Right ascension in decimal degrees (ICRS)\\
\texttt{DEdeg} & deg & Declination in decimal degrees (ICRS)\\
\texttt{Grade} & --- & Visual inspection grade for conventional lenses\\
\texttt{z(d)} & --- & Lens spectroscopic redshift\\
\texttt{z(s)} & --- & Source spectroscopic redshift\\
\texttt{Known} & --- & Previously reported lens candidate\\
\texttt{Dimple} & --- & Dimple candidate\\
\texttt{DGC} & --- & Matched with DGC\tablenotemark{a} \\
\texttt{Sep} & arcsec & Angular separation between the fiber pair\\
\texttt{RE} & arcsec & Estimated Einstein radius\\
\texttt{E\_RE} & arcsec & $+1\sigma$ limit of the estimated Einstein radius\\
\texttt{e\_RE} & arcsec & $-1\sigma$ limit of the estimated Einstein radius\\
\texttt{Vd} & km/s & Lens velocity dispersion\\
\texttt{e\_Vd} & km/s & Uncertainty of the lens velocity dispersion\\
\texttt{NSp} & --- & Number of spectra with reliable redshift measurements in the group\\
\texttt{Nz} & --- & Number of distinct redshifts ($\Delta z\ge0.02$) in the group\\
\texttt{Nz1} & --- & Number of spectra with the first redshift in the group\\
\texttt{Nz2} & --- & Number of spectra with the second redshift in the group\\
\texttt{Nz3} & --- & Number of spectra with the third redshift in the group\\
\texttt{q\_Sp(d)} & --- & Lens spectroscopic quality\\
\texttt{q\_Sp(s)} & --- & Source spectroscopic quality\\
\texttt{SpType(d)} & --- & Lens spectral type\\
\texttt{SpType(s)} & --- & Source spectral type\\
\texttt{Emi(d)} & --- & Emission lines present in the lens spectrum\tablenotemark{b} \\
\texttt{Emi(s)} & --- & Emission lines present in the source spectrum\tablenotemark{b} \\
\texttt{Abs(d)} & --- & Absorption lines present in the lens spectrum\tablenotemark{b} \\
\texttt{Abs(s)} & --- & Absorption lines present in the source spectrum\tablenotemark{b} \\
\texttt{Fg(d)} & nanomaggy & Lens $g$-band flux from the tractor (FLUX\_G)\\
\texttt{Fr(d)} & nanomaggy & Lens $r$-band flux from the tractor (FLUX\_R)\\
\texttt{Fz(d)} & nanomaggy & Lens $z$-band flux from the tractor (FLUX\_Z)\\
\texttt{Fg(s)} & nanomaggy & Source $g$-band flux from the tractor (FLUX\_G)\\
\texttt{Fr(s)} & nanomaggy & Source $r$-band flux from the tractor (FLUX\_R)\\
\texttt{Fz(s)} & nanomaggy & Source $z$-band flux from the tractor (FLUX\_Z)\\
\texttt{Tid(d)} & --- & Lens \texttt{TARGETID} in the DESI catalog\\
\texttt{Sur(d)} & --- & Lens \texttt{SURVEY} name in the DESI catalog\\
\texttt{Prog(d)} & --- & Lens \texttt{PROGRAM} name in the DESI catalog\\
\texttt{Tid(s)} & --- & Source \texttt{TARGETID} in the DESI catalog\\
\texttt{Sur(s)} & --- & Source \texttt{SURVEY} name in the DESI catalog\\
\texttt{Prog(s)} & --- & Source \texttt{PROGRAM} name in the DESI catalog\\
\enddata
\tablenotetext{a}{Available only for the dimple candidates.}
\tablenotetext{b}{Available only for the spectral type of galaxy.}
\tablecomments{(\texttt{TARGETID}, \texttt{SURVEY}, \texttt{PROGRAM}) can uniquely determine a fiber in the DESI redshift catalogs.}
\end{deluxetable*}

\begin{figure*}
    \centering
    \includegraphics[width=\linewidth]{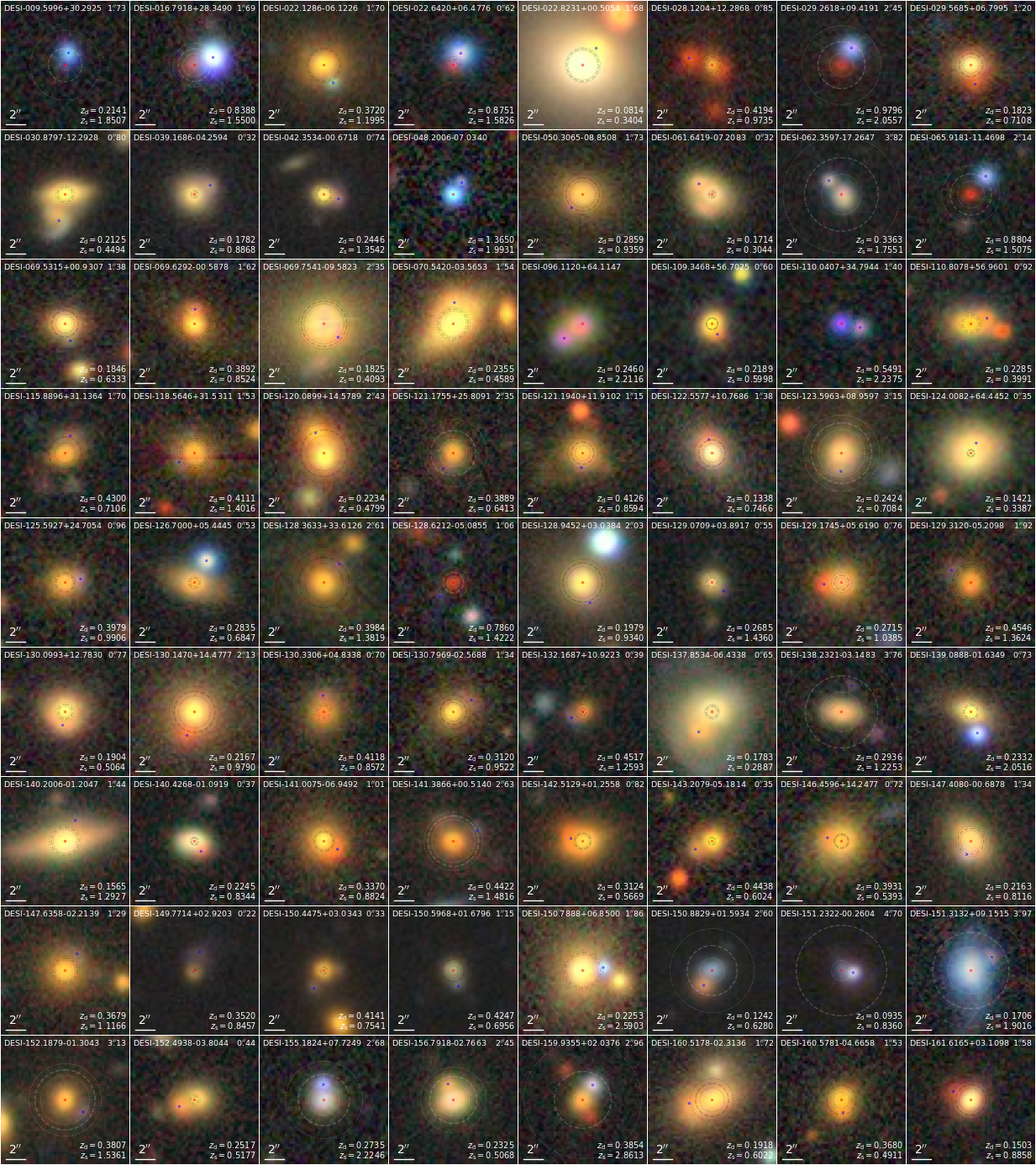}
    \caption{All new Grade B candidates discovered in this work. The formatting and annotations are the same as in Figure~\ref{fig:newA}.}
    \label{fig:newB}
\end{figure*}

\begin{figure*}
    \centering
    \includegraphics[width=\linewidth]{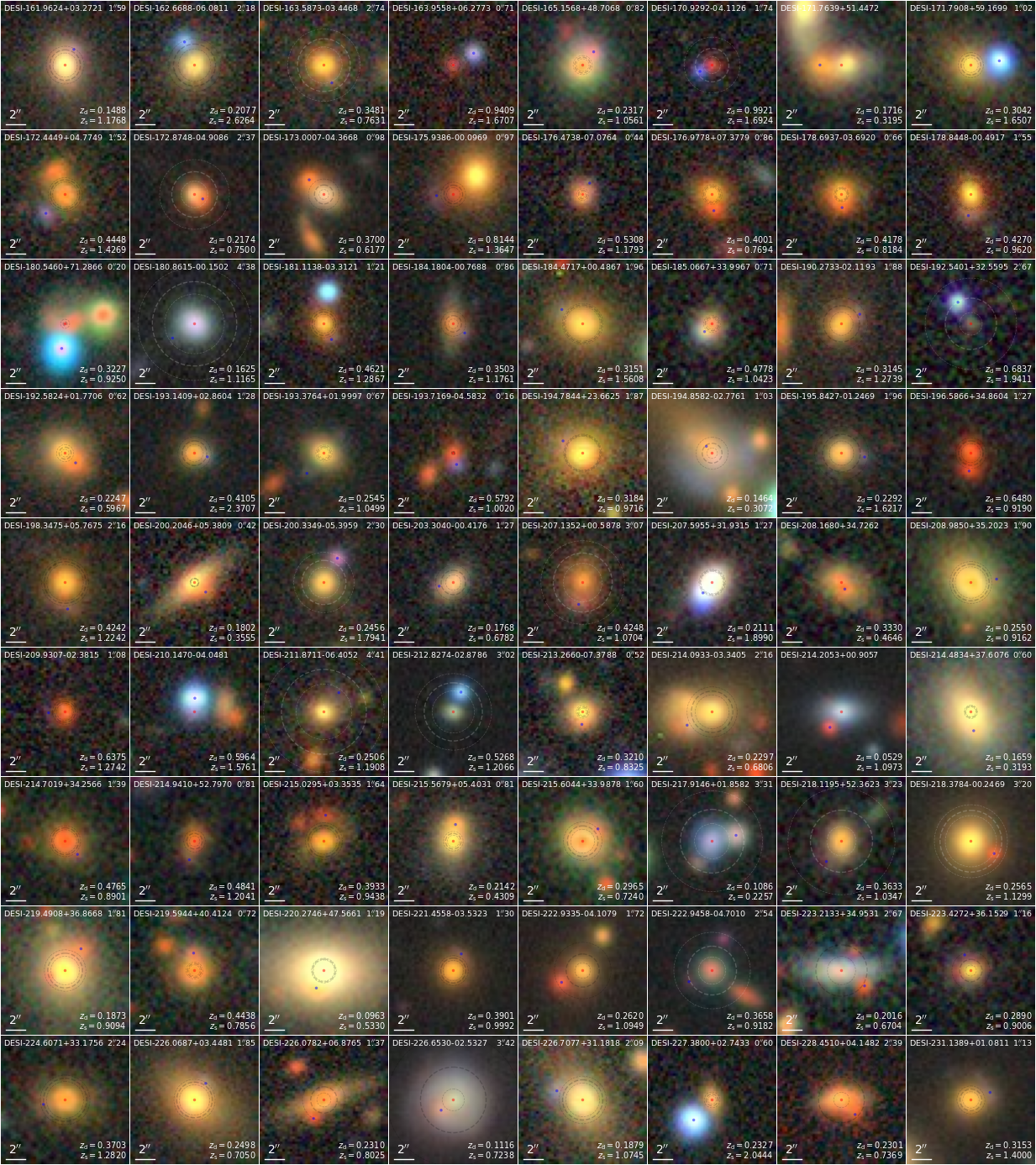}
    \caption{(Continued)}
\end{figure*}

\begin{figure*}
    \centering
    \includegraphics[width=\linewidth]{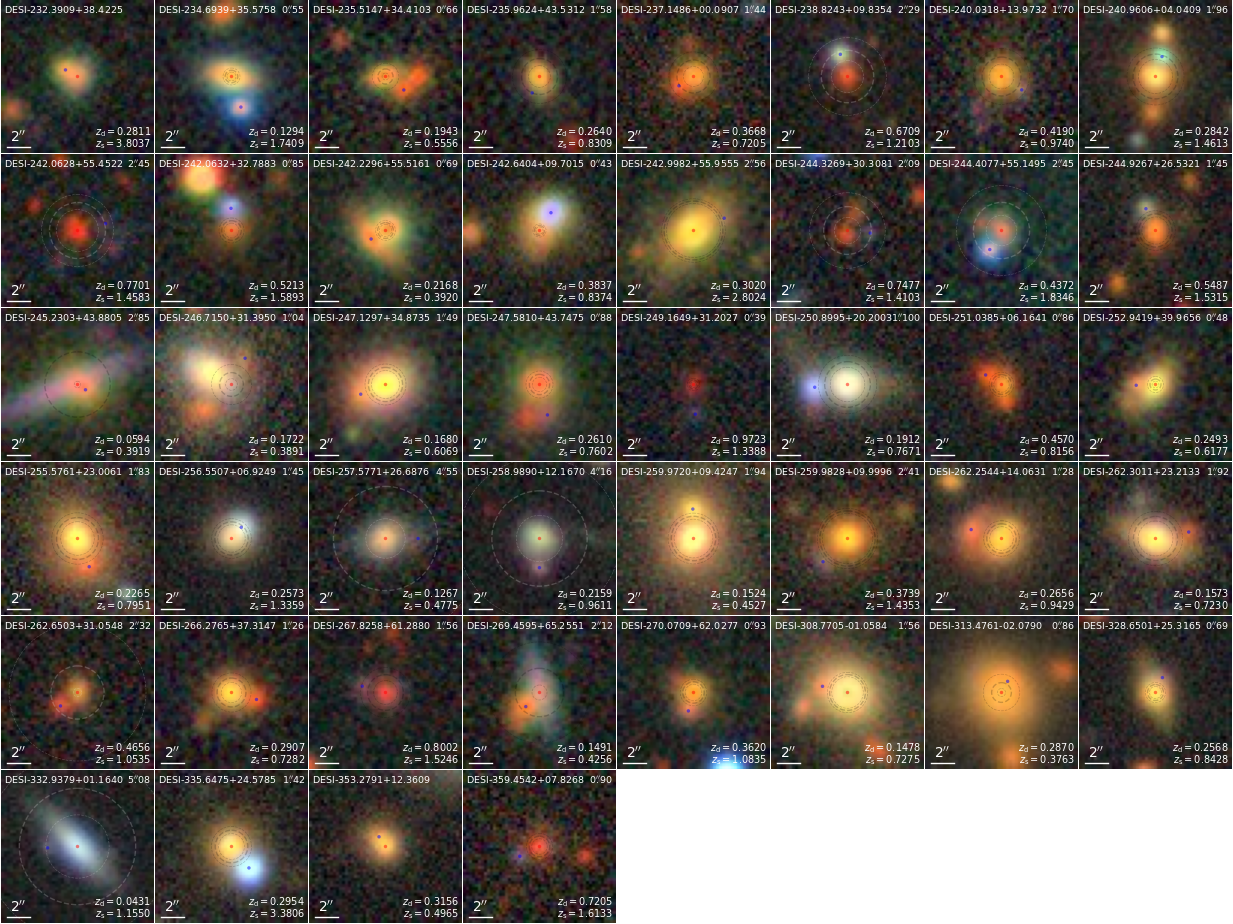}
    \caption{(Continued)}
\end{figure*}

\begin{figure*}
    \centering
    \includegraphics[width=\linewidth]{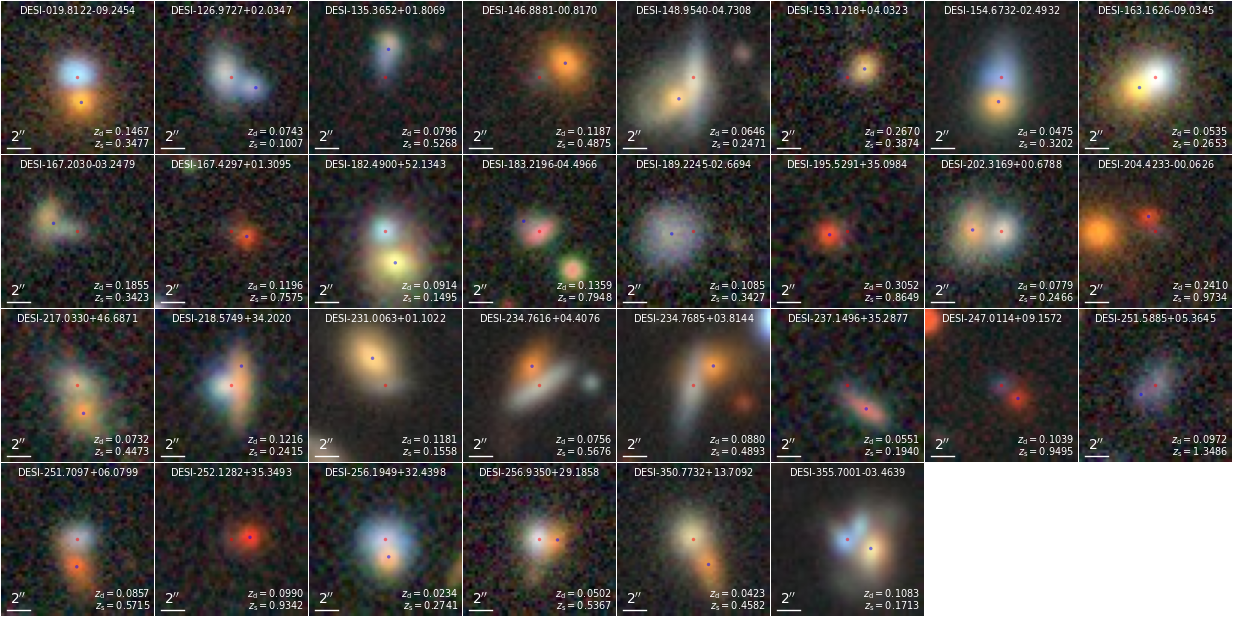}
    \caption{All 30 dimple candidates matched with the dwarf galaxies in the DGC. The formatting and annotations are the same as in Figure~\ref{fig:newA}. The system DESI-189.2245-02.6694 is also shown in Figure~\ref{fig:dimple_exp}.}
    \label{fig:dwarf}
\end{figure*}

\bibliography{references}{}

\begin{thebibliography}{}
\expandafter\ifx\csname natexlab\endcsname\relax\def\natexlab#1{#1}\fi
\providecommand{\url}[1]{\href{#1}{#1}}
\providecommand{\dodoi}[1]{doi:~\href{http://doi.org/#1}{\nolinkurl{#1}}}
\providecommand{\doeprint}[1]{\href{http://ascl.net/#1}{\nolinkurl{http://ascl.net/#1}}}
\providecommand{\doarXiv}[1]{\href{https://arxiv.org/abs/#1}{\nolinkurl{https://arxiv.org/abs/#1}}}

\bibitem[{A. {Amruth} {et~al.}(2023){Amruth}, {Broadhurst}, {Lim}, {Oguri}, {Smoot}, {Diego}, {Leung}, {Emami}, {Li}, {Chiueh}, {Schive}, {Yeung}, \& {Li}}]{Amruth2023}
{Amruth}, A., {Broadhurst}, T., {Lim}, J., {et~al.} 2023, \bibinfo{title}{{Einstein rings modulated by wavelike dark matter from anomalies in gravitationally lensed images},} Nature Astronomy, 7, 736, \dodoi{10.1038/s41550-023-01943-9}

\bibitem[{A. {Anand} {et~al.}(2024){Anand}, {Guy}, {Bailey}, {Moustakas}, {Aguilar}, {Ahlen}, {Bolton}, {Brodzeller}, {Brooks}, {Claybaugh}, {Cole}, {de la Macorra}, {Dey}, {Fanning}, {Forero-Romero}, {Gazta{\~n}aga}, {Gontcho A Gontcho}, {Gutierrez}, {Honscheid}, {Howlett}, {Juneau}, {Kirkby}, {Kisner}, {Kremin}, {Lambert}, {Landriau}, {Le Guillou}, {Manera}, {Meisner}, {Miquel}, {Mueller}, {Niz}, {Palanque-Delabrouille}, {Percival}, {Poppett}, {Prada}, {Raichoor}, {Rezaie}, {Rossi}, {Sanchez}, {Schlafly}, {Schlegel}, {Schubnell}, {Sprayberry}, {Tarl{\'e}}, {Warner}, {Weaver}, {Zhou}, \& {Zou}}]{Anand2024}
{Anand}, A., {Guy}, J., {Bailey}, S., {et~al.} 2024, \bibinfo{title}{{Archetype-based Redshift Estimation for the Dark Energy Spectroscopic Instrument Survey},} \aj, 168, 124, \dodoi{10.3847/1538-3881/ad60c2}

\bibitem[{ {Astropy Collaboration} {et~al.}(2013){Astropy Collaboration}, {Robitaille}, {Tollerud}, {Greenfield}, {Droettboom}, {Bray}, {Aldcroft}, {Davis}, {Ginsburg}, {Price-Whelan}, {Kerzendorf}, {Conley}, {Crighton}, {Barbary}, {Muna}, {Ferguson}, {Grollier}, {Parikh}, {Nair}, {Unther}, {Deil}, {Woillez}, {Conseil}, {Kramer}, {Turner}, {Singer}, {Fox}, {Weaver}, {Zabalza}, {Edwards}, {Azalee Bostroem}, {Burke}, {Casey}, {Crawford}, {Dencheva}, {Ely}, {Jenness}, {Labrie}, {Lim}, {Pierfederici}, {Pontzen}, {Ptak}, {Refsdal}, {Servillat}, \& {Streicher}}]{AstropyCollaboration2013}
{Astropy Collaboration}, {Robitaille}, T.~P., {Tollerud}, E.~J., {et~al.} 2013, \bibinfo{title}{{Astropy: A community Python package for astronomy},} \aap, 558, A33, \dodoi{10.1051/0004-6361/201322068}

\bibitem[{ {Astropy Collaboration} {et~al.}(2018){Astropy Collaboration}, {Price-Whelan}, {Sip{\H{o}}cz}, {G{\"u}nther}, {Lim}, {Crawford}, {Conseil}, {Shupe}, {Craig}, {Dencheva}, \& et~al.}]{AstropyCollaboration2018}
{Astropy Collaboration}, {Price-Whelan}, A.~M., {Sip{\H{o}}cz}, B.~M., {et~al.} 2018, \bibinfo{title}{{The Astropy Project: Building an Open-science Project and Status of the v2.0 Core Package},} \aj, 156, 123, \dodoi{10.3847/1538-3881/aabc4f}

\bibitem[{ {Astropy Collaboration} {et~al.}(2022){Astropy Collaboration}, {Price-Whelan}, {Lim}, {Earl}, {Starkman}, {Bradley}, {Shupe}, {Patil}, {Corrales}, {Brasseur}, \& et~al.}]{AstropyCollaboration2022}
{Astropy Collaboration}, {Price-Whelan}, A.~M., {Lim}, P.~L., {et~al.} 2022, \bibinfo{title}{{The Astropy Project: Sustaining and Growing a Community-oriented Open-source Project and the Latest Major Release (v5.0) of the Core Package},} \apj, 935, 167, \dodoi{10.3847/1538-4357/ac7c74}

\bibitem[{A.~S. {Bolton} {et~al.}(2008){Bolton}, {Burles}, {Koopmans}, {Treu}, {Gavazzi}, {Moustakas}, {Wayth}, \& {Schlegel}}]{Bolton2008}
{Bolton}, A.~S., {Burles}, S., {Koopmans}, L. V.~E., {et~al.} 2008, \bibinfo{title}{{The Sloan Lens ACS Survey. V. The Full ACS Strong-Lens Sample},} \apj, 682, 964, \dodoi{10.1086/589327}

\bibitem[{A.~S. {Bolton} {et~al.}(2006){Bolton}, {Burles}, {Koopmans}, {Treu}, \& {Moustakas}}]{Bolton2006}
{Bolton}, A.~S., {Burles}, S., {Koopmans}, L. V.~E., {Treu}, T., \& {Moustakas}, L.~A. 2006, \bibinfo{title}{{The Sloan Lens ACS Survey. I. A Large Spectroscopically Selected Sample of Massive Early-Type Lens Galaxies},} \apj, 638, 703, \dodoi{10.1086/498884}

\bibitem[{A. {Brodzeller} {et~al.}(2023){Brodzeller}, {Dawson}, {Bailey}, {Yu}, {Ross}, {Bault}, {Filbert}, {Aguilar}, {Ahlen}, {Alexander}, {Armengaud}, {Berti}, {Brooks}, {Chaussidon}, {de la Macorra}, {Doel}, {Fanning}, {Fawcett}, {Font-Ribera}, {A Gontcho}, {Guy}, {Honscheid}, {Juneau}, {Kehoe}, {Kisner}, {Kremin}, {Lan}, {Landriau}, {Levi}, {Magneville}, {Martini}, {Meisner}, {Miquel}, {Moustakas}, {Palanque-Delabrouille}, {Percival}, {Prada}, {Ravoux}, {Rossi}, {Saulder}, {Siudek}, {Tarl{\'e}}, {Weaver}, {Youles}, {Zheng}, {Zhou}, \& {Zhou}}]{Brodzeller2023}
{Brodzeller}, A., {Dawson}, K., {Bailey}, S., {et~al.} 2023, \bibinfo{title}{{Performance of the Quasar Spectral Templates for the Dark Energy Spectroscopic Instrument},} \aj, 166, 66, \dodoi{10.3847/1538-3881/ace35d}

\bibitem[{J.~R. {Brownstein} {et~al.}(2012){Brownstein}, {Bolton}, {Schlegel}, {Eisenstein}, {Kochanek}, {Connolly}, {Maraston}, {Pandey}, {Seitz}, {Wake}, {Wood-Vasey}, {Brinkmann}, {Schneider}, \& {Weaver}}]{Brownstein2012}
{Brownstein}, J.~R., {Bolton}, A.~S., {Schlegel}, D.~J., {et~al.} 2012, \bibinfo{title}{{The BOSS Emission-Line Lens Survey (BELLS). I. A Large Spectroscopically Selected Sample of Lens Galaxies at Redshift \raisebox{-0.5ex}\textasciitilde0.5},} \apj, 744, 41, \dodoi{10.1088/0004-637X/744/1/41}

\bibitem[{R. {Ca{\~n}ameras} {et~al.}(2020){Ca{\~n}ameras}, {Schuldt}, {Suyu}, {Taubenberger}, {Meinhardt}, {Leal-Taix{\'e}}, {Lemon}, {Rojas}, \& {Savary}}]{Canameras2020}
{Ca{\~n}ameras}, R., {Schuldt}, S., {Suyu}, S.~H., {et~al.} 2020, \bibinfo{title}{{HOLISMOKES. II. Identifying galaxy-scale strong gravitational lenses in Pan-STARRS using convolutional neural networks},} \aap, 644, A163, \dodoi{10.1051/0004-6361/202038219}

\bibitem[{R. {Ca{\~n}ameras} {et~al.}(2021){Ca{\~n}ameras}, {Schuldt}, {Shu}, {Suyu}, {Taubenberger}, {Meinhardt}, {Leal-Taix{\'e}}, {Chao}, {Inoue}, {Jaelani}, \& {More}}]{Canameras2021}
{Ca{\~n}ameras}, R., {Schuldt}, S., {Shu}, Y., {et~al.} 2021, \bibinfo{title}{{HOLISMOKES. VI. New galaxy-scale strong lens candidates from the HSC-SSP imaging survey},} \aap, 653, L6, \dodoi{10.1051/0004-6361/202141758}

\bibitem[{J.~H.~H. {Chan} {et~al.}(2020{\natexlab{a}}){Chan}, {Schive}, {Wong}, {Chiueh}, \& {Broadhurst}}]{Chan2020a}
{Chan}, J. H.~H., {Schive}, H.-Y., {Wong}, S.-K., {Chiueh}, T., \& {Broadhurst}, T. 2020{\natexlab{a}}, \bibinfo{title}{{Multiple Images and Flux Ratio Anomaly of Fuzzy Gravitational Lenses},} \prl, 125, 111102, \dodoi{10.1103/PhysRevLett.125.111102}

\bibitem[{J.~H.~H. {Chan} {et~al.}(2020{\natexlab{b}}){Chan}, {Suyu}, {Sonnenfeld}, {Jaelani}, {More}, {Yonehara}, {Kubota}, {Coupon}, {Lee}, {Oguri}, {Rusu}, \& {Wong}}]{Chan2020b}
{Chan}, J. H.~H., {Suyu}, S.~H., {Sonnenfeld}, A., {et~al.} 2020{\natexlab{b}}, \bibinfo{title}{{Survey of Gravitationally lensed Objects in HSC Imaging (SuGOHI). IV. Lensed quasar search in the HSC survey},} \aap, 636, A87, \dodoi{10.1051/0004-6361/201937030}

\bibitem[{E. {Chaussidon} {et~al.}(2023){Chaussidon}, {Y{\`e}che}, {Palanque-Delabrouille}, {Alexander}, {Yang}, {Ahlen}, {Bailey}, {Brooks}, {Cai}, {Chabanier}, {Davis}, {Dawson}, {de laMacorra}, {Dey}, {Dey}, {Eftekharzadeh}, {Eisenstein}, {Fanning}, {Font-Ribera}, {Gazta{\~n}aga}, {A Gontcho}, {Gonzalez-Morales}, {Guy}, {Herrera-Alcantar}, {Honscheid}, {Ishak}, {Jiang}, {Juneau}, {Kehoe}, {Kisner}, {Kov{\'a}cs}, {Kremin}, {Lan}, {Landriau}, {Le Guillou}, {Levi}, {Magneville}, {Martini}, {Meisner}, {Moustakas}, {Mu{\~n}oz-Guti{\'e}rrez}, {Myers}, {Newman}, {Nie}, {Percival}, {Poppett}, {Prada}, {Raichoor}, {Ravoux}, {Ross}, {Schlafly}, {Schlegel}, {Tan}, {Tarl{\'e}}, {Zhou}, {Zhou}, \& {Zou}}]{Chaussidon2023}
{Chaussidon}, E., {Y{\`e}che}, C., {Palanque-Delabrouille}, N., {et~al.} 2023, \bibinfo{title}{{Target Selection and Validation of DESI Quasars},} \apj, 944, 107, \dodoi{10.3847/1538-4357/acb3c2}

\bibitem[{ {DESI Collaboration} {et~al.}(2016{\natexlab{a}}){DESI Collaboration}, {Aghamousa}, {Aguilar}, {Ahlen}, {Alam}, {Allen}, {Allende Prieto}, {Annis}, {Bailey}, {Balland}, \& et~al.}]{DESICollaboration2016a}
{DESI Collaboration}, {Aghamousa}, A., {Aguilar}, J., {et~al.} 2016{\natexlab{a}}, \bibinfo{title}{{The DESI Experiment Part I: Science,Targeting, and Survey Design},} arXiv e-prints, arXiv:1611.00036, \dodoi{10.48550/arXiv.1611.00036}

\bibitem[{ {DESI Collaboration} {et~al.}(2016{\natexlab{b}}){DESI Collaboration}, {Aghamousa}, {Aguilar}, {Ahlen}, {Alam}, {Allen}, {Allende Prieto}, {Annis}, {Bailey}, {Balland}, \& et~al.}]{DESICollaboration2016b}
{DESI Collaboration}, {Aghamousa}, A., {Aguilar}, J., {et~al.} 2016{\natexlab{b}}, \bibinfo{title}{{The DESI Experiment Part II: Instrument Design},} arXiv e-prints, arXiv:1611.00037, \dodoi{10.48550/arXiv.1611.00037}

\bibitem[{ {DESI Collaboration} {et~al.}(2022){DESI Collaboration}, {Abareshi}, {Aguilar}, {Ahlen}, {Alam}, {Alexander}, {Alfarsy}, {Allen}, {Allende Prieto}, {Alves}, \& et~al.}]{DESICollaboration2022}
{DESI Collaboration}, {Abareshi}, B., {Aguilar}, J., {et~al.} 2022, \bibinfo{title}{{Overview of the Instrumentation for the Dark Energy Spectroscopic Instrument},} \aj, 164, 207, \dodoi{10.3847/1538-3881/ac882b}

\bibitem[{ {DESI Collaboration} {et~al.}(2025){DESI Collaboration}, {Abdul-Karim}, {Adame}, {Aguado}, {Aguilar}, {Ahlen}, {Alam}, {Aldering}, {Alexander}, {Alfarsy}, \& et~al.}]{DESICollaboration2025}
{DESI Collaboration}, {Abdul-Karim}, M., {Adame}, A.~G., {et~al.} 2025, \bibinfo{title}{{Data Release 1 of the Dark Energy Spectroscopic Instrument},} arXiv e-prints, arXiv:2503.14745, \dodoi{10.48550/arXiv.2503.14745}

\bibitem[{A. {Dey} {et~al.}(2019){Dey}, {Schlegel}, {Lang}, {Blum}, {Burleigh}, {Fan}, {Findlay}, {Finkbeiner}, {Herrera}, {Juneau}, \& et~al.}]{Dey2019}
{Dey}, A., {Schlegel}, D.~J., {Lang}, D., {et~al.} 2019, \bibinfo{title}{{Overview of the DESI Legacy Imaging Surveys},} \aj, 157, 168, \dodoi{10.3847/1538-3881/ab089d}

\bibitem[{S.~P. {Driver} {et~al.}(2022){Driver}, {Robotham}, {Obreschkow}, {Peacock}, {Baldry}, {Bellstedt}, {Bland-Hawthorn}, {Brough}, {Cluver}, {Holwerda}, {Hopkins}, {Lagos}, {Liske}, {Loveday}, {Phillipps}, \& {Taylor}}]{Driver2022}
{Driver}, S.~P., {Robotham}, A. S.~G., {Obreschkow}, D., {et~al.} 2022, \bibinfo{title}{{An empirical measurement of the halo mass function from the combination of GAMA DR4, SDSS DR12, and REFLEX II data},} \mnras, 515, 2138, \dodoi{10.1093/mnras/stac581}

\bibitem[{E.~O. {Garvin} {et~al.}(2022){Garvin}, {Kruk}, {Cornen}, {Bhatawdekar}, {Ca{\~n}ameras}, \& {Mer{\'\i}n}}]{Garvin2022}
{Garvin}, E.~O., {Kruk}, S., {Cornen}, C., {et~al.} 2022, \bibinfo{title}{{Hubble Asteroid Hunter. II. Identifying strong gravitational lenses in HST images with crowdsourcing},} \aap, 667, A141, \dodoi{10.1051/0004-6361/202243745}

\bibitem[{R. {Gavazzi} {et~al.}(2014){Gavazzi}, {Marshall}, {Treu}, \& {Sonnenfeld}}]{Gavazzi2014}
{Gavazzi}, R., {Marshall}, P.~J., {Treu}, T., \& {Sonnenfeld}, A. 2014, \bibinfo{title}{{RINGFINDER: Automated Detection of Galaxy-scale Gravitational Lenses in Ground-based Multi-filter Imaging Data},} \apj, 785, 144, \dodoi{10.1088/0004-637X/785/2/144}

\bibitem[{J. {Gonz{\'a}lez-Nuevo} {et~al.}(2019){Gonz{\'a}lez-Nuevo}, {Su{\'a}rez G{\'o}mez}, {Bonavera}, {S{\'a}nchez-Lasheras}, {Arg{\"u}eso}, {Toffolatti}, {Herranz}, {Gonz{\'a}lez-Guti{\'e}rrez}, {Garc{\'\i}a Riesgo}, \& {de Cos Juez}}]{Gonzalez-Nuevo2019}
{Gonz{\'a}lez-Nuevo}, J., {Su{\'a}rez G{\'o}mez}, S.~L., {Bonavera}, L., {et~al.} 2019, \bibinfo{title}{{SHALOS: Statistical Herschel-ATLAS lensed objects selection},} \aap, 627, A31, \dodoi{10.1051/0004-6361/201935475}

\bibitem[{J. {Guy} {et~al.}(2023){Guy}, {Bailey}, {Kremin}, {Alam}, {Alexander}, {Allende Prieto}, {BenZvi}, {Bolton}, {Brooks}, {Chaussidon}, {Cooper}, {Dawson}, {de la Macorra}, {Dey}, {Dey}, {Dhungana}, {Eisenstein}, {Font-Ribera}, {Forero-Romero}, {Gazta{\~n}aga}, {Gontcho A Gontcho}, {Green}, {Honscheid}, {Ishak}, {Kehoe}, {Kirkby}, {Kisner}, {Koposov}, {Lan}, {Landriau}, {Le Guillou}, {Levi}, {Magneville}, {Manser}, {Martini}, {Meisner}, {Miquel}, {Moustakas}, {Myers}, {Newman}, {Nie}, {Palanque-Delabrouille}, {Percival}, {Poppett}, {Prada}, {Raichoor}, {Ravoux}, {Ross}, {Schlafly}, {Schlegel}, {Schubnell}, {Sharples}, {Tarl{\'e}}, {Weaver}, {Y{\'e}che}, {Zhou}, {Zhou}, \& {Zou}}]{Guy2023}
{Guy}, J., {Bailey}, S., {Kremin}, A., {et~al.} 2023, \bibinfo{title}{{The Spectroscopic Data Processing Pipeline for the Dark Energy Spectroscopic Instrument},} \aj, 165, 144, \dodoi{10.3847/1538-3881/acb212}

\bibitem[{C. {Hahn} {et~al.}(2023){Hahn}, {Wilson}, {Ruiz-Macias}, {Cole}, {Weinberg}, {Moustakas}, {Kremin}, {Tinker}, {Smith}, {Wechsler}, {Ahlen}, {Alam}, {Bailey}, {Brooks}, {Cooper}, {Davis}, {Dawson}, {Dey}, {Dey}, {Eftekharzadeh}, {Eisenstein}, {Fanning}, {Forero-Romero}, {Frenk}, {Gazta{\~n}aga}, {A Gontcho}, {Guy}, {Honscheid}, {Ishak}, {Juneau}, {Kehoe}, {Kisner}, {Lan}, {Landriau}, {Le Guillou}, {Levi}, {Magneville}, {Martini}, {Meisner}, {Myers}, {Nie}, {Norberg}, {Palanque-Delabrouille}, {Percival}, {Poppett}, {Prada}, {Raichoor}, {Ross}, {Gaines}, {Saulder}, {Schlafly}, {Schlegel}, {Sierra-Porta}, {Tarle}, {Weaver}, {Y{\`e}che}, {Zarrouk}, {Zhou}, {Zhou}, \& {Zou}}]{Hahn2023}
{Hahn}, C., {Wilson}, M.~J., {Ruiz-Macias}, O., {et~al.} 2023, \bibinfo{title}{{The DESI Bright Galaxy Survey: Final Target Selection, Design, and Validation},} \aj, 165, 253, \dodoi{10.3847/1538-3881/accff8}

\bibitem[{Q. {He} {et~al.}(2025){He}, {Robertson}, {Nightingale}, {Amvrosiadis}, {Cole}, {Frenk}, {Lange}, {Li}, {Li}, {Cao}, {Fung}, {Ma}, {Massey}, {Wang}, \& {von Wietersheim-Kramsta}}]{He2025}
{He}, Q., {Robertson}, A., {Nightingale}, J.~W., {et~al.} 2025, \bibinfo{title}{{Not so dark, not so dense: an alternative explanation for the lensing subhalo in SDSSJ0946+1006},} arXiv e-prints, arXiv:2506.07978, \dodoi{10.48550/arXiv.2506.07978}

\bibitem[{Y.-M. {Hsu}(2025){Hsu}}]{Hsu2025}
{Hsu}, Y.-M. 2025, \bibinfo{title}{{spherimatch: Cross-matching and self-matching in spherical coordinates},}, Astrophysics Source Code Library, record ascl:2507.022

\bibitem[{X. {Huang} {et~al.}(2020){Huang}, {Storfer}, {Ravi}, {Pilon}, {Domingo}, {Schlegel}, {Bailey}, {Dey}, {Gupta}, {Herrera}, {Juneau}, {Landriau}, {Lang}, {Meisner}, {Moustakas}, {Myers}, {Schlafly}, {Valdes}, {Weaver}, {Yang}, \& {Y{\`e}che}}]{Huang2020}
{Huang}, X., {Storfer}, C., {Ravi}, V., {et~al.} 2020, \bibinfo{title}{{Finding Strong Gravitational Lenses in the DESI DECam Legacy Survey},} \apj, 894, 78, \dodoi{10.3847/1538-4357/ab7ffb}

\bibitem[{X. {Huang} {et~al.}(2021){Huang}, {Storfer}, {Gu}, {Ravi}, {Pilon}, {Sheu}, {Venguswamy}, {Banka}, {Dey}, {Landriau}, {Lang}, {Meisner}, {Moustakas}, {Myers}, {Sajith}, {Schlafly}, \& {Schlegel}}]{Huang2021}
{Huang}, X., {Storfer}, C., {Gu}, A., {et~al.} 2021, \bibinfo{title}{{Discovering New Strong Gravitational Lenses in the DESI Legacy Imaging Surveys},} \apj, 909, 27, \dodoi{10.3847/1538-4357/abd62b}

\bibitem[{X. {Huang} {et~al.}(2025){Huang}, {Baltasar}, {Ratier-Werbin}, {Storfer}, {Sheu}, {Agarwal}, {Tamargo-Arizmendi}, {Schlegel}, {Aguilar}, {Ahlen}, {Aldering}, {Banka}, {BenZvi}, {Bianchi}, {Bolton}, {Brooks}, {Cikota}, {Claybaugh}, {de la Macorra}, {Dey}, {Doel}, {Edelstein}, {Filipp}, {Forero-Romero}, {Gaztanaga}, {Gontcho}, {Gu}, {Gutierrez}, {Honscheid}, {Jullo}, {Juneau}, {Kehoe}, {Kirkby}, {Kisner}, {Kremin}, {Kwon}, {Lambert}, {Landriau}, {Lang}, {Le Guillou}, {Liu}, {Meisner}, {Miquel}, {Moustakas}, {Myers}, {Perlmutter}, {Perez-Rafols}, {Prada}, {Rossi}, {Rubin}, {Sanchez}, {Schubnell}, {Shu}, {Silver}, {Sprayberry}, {Suzuki}, {Tarle}, {Weaver}, \& {Zou}}]{Huang2025}
{Huang}, X., {Baltasar}, S., {Ratier-Werbin}, N., {et~al.} 2025, \bibinfo{title}{{DESI Strong Lens Foundry I: HST Observations and Modeling with GIGA-Lens},} arXiv e-prints, arXiv:2502.03455, \dodoi{10.48550/arXiv.2502.03455}

\bibitem[{J.~C. {Inchausti} {et~al.}(2025){Inchausti}, {Storfer}, {Huang}, {Hsu}, {Kaufmann}, {Pasupala}, {Banka}, {Dey}, {Lang}, {Meisner}, {Moustakas}, {Myers}, {Schlafly}, \& {Schlegel}}]{Inchausti2025}
{Inchausti}, J.~C., {Storfer}, C.~J., {Huang}, X., {et~al.} 2025, \bibinfo{title}{{Strong Lens Discoveries in DESI Legacy Imaging Surveys DR10 with Two Deep Learning Architectures},} arXiv e-prints, arXiv:2508.20087, \dodoi{10.48550/arXiv.2508.20087}

\bibitem[{K.~T. {Inoue} {et~al.}(2024){Inoue}, {Shinohara}, {Suyama}, \& {Takahashi}}]{Inoue2024}
{Inoue}, K.~T., {Shinohara}, T., {Suyama}, T., \& {Takahashi}, T. 2024, \bibinfo{title}{{Probing warm and mixed dark matter models using lensing shift power spectrum},} \prd, 109, 103509, \dodoi{10.1103/PhysRevD.109.103509}

\bibitem[{C. {Jacobs} {et~al.}(2019){Jacobs}, {Collett}, {Glazebrook}, {Buckley-Geer}, {Diehl}, {Lin}, {McCarthy}, {Qin}, {Odden}, {Caso Escudero}, {Dial}, {Yung}, {Gaitsch}, {Pellico}, {Lindgren}, {Abbott}, {Annis}, {Avila}, {Brooks}, {Burke}, {Carnero Rosell}, {Carrasco Kind}, {Carretero}, {da Costa}, {De Vicente}, {Fosalba}, {Frieman}, {Garc{\'\i}a-Bellido}, {Gaztanaga}, {Goldstein}, {Gruen}, {Gruendl}, {Gschwend}, {Hollowood}, {Honscheid}, {Hoyle}, {James}, {Krause}, {Kuropatkin}, {Lahav}, {Lima}, {Maia}, {Marshall}, {Miquel}, {Plazas}, {Roodman}, {Sanchez}, {Scarpine}, {Serrano}, {Sevilla-Noarbe}, {Smith}, {Sobreira}, {Suchyta}, {Swanson}, {Tarle}, {Vikram}, {Walker}, {Zhang}, \& {DES Collaboration}}]{Jacobs2019}
{Jacobs}, C., {Collett}, T., {Glazebrook}, K., {et~al.} 2019, \bibinfo{title}{{An Extended Catalog of Galaxy-Galaxy Strong Gravitational Lenses Discovered in DES Using Convolutional Neural Networks},} \apjs, 243, 17, \dodoi{10.3847/1538-4365/ab26b6}

\bibitem[{A.~T. {Jaelani} {et~al.}(2020){Jaelani}, {More}, {Oguri}, {Sonnenfeld}, {Suyu}, {Rusu}, {Wong}, {Chan}, {Kayo}, {Lee}, {Chao}, {Coupon}, {Inoue}, \& {Futamase}}]{Jaelani2020}
{Jaelani}, A.~T., {More}, A., {Oguri}, M., {et~al.} 2020, \bibinfo{title}{{Survey of Gravitationally lensed Objects in HSC Imaging (SuGOHI) - V. Group-to-cluster scale lens search from the HSC-SSP Survey},} \mnras, 495, 1291, \dodoi{10.1093/mnras/staa1062}

\bibitem[{A. {Jenkins} {et~al.}(2001){Jenkins}, {Frenk}, {White}, {Colberg}, {Cole}, {Evrard}, {Couchman}, \& {Yoshida}}]{Jenkins2001}
{Jenkins}, A., {Frenk}, C.~S., {White}, S.~D.~M., {et~al.} 2001, \bibinfo{title}{{The mass function of dark matter haloes},} \mnras, 321, 372, \dodoi{10.1046/j.1365-8711.2001.04029.x}

\bibitem[{B.~C. {Lacki} {et~al.}(2009){Lacki}, {Kochanek}, {Stanek}, {Inada}, \& {Oguri}}]{Lacki2009}
{Lacki}, B.~C., {Kochanek}, C.~S., {Stanek}, K.~Z., {Inada}, N., \& {Oguri}, M. 2009, \bibinfo{title}{{Difference Imaging of Lensed Quasar Candidates in the Sloan Digital Sky Survey Supernova Survey Region},} \apj, 698, 428, \dodoi{10.1088/0004-637X/698/1/428}

\bibitem[{T.-W. {Lan} {et~al.}(2023){Lan}, {Tojeiro}, {Armengaud}, {Prochaska}, {Davis}, {Alexander}, {Raichoor}, {Zhou}, {Y{\`e}che}, {Balland}, {BenZvi}, {Berti}, {Canning}, {Carr}, {Chittenden}, {Cole}, {Cousinou}, {Dawson}, {Dey}, {Douglass}, {Edge}, {Escoffier}, {Glanville}, {A Gontcho}, {Guy}, {Hahn}, {Howlett}, {Hwang}, {Jiang}, {Kov{\'a}cs}, {Mezcua}, {Moore}, {Nadathur}, {Oh}, {Parkinson}, {Rocher}, {Ross}, {Ruhlmann-Kleider}, {Sabiu}, {Said}, {Saulder}, {Sierra-Porta}, {Weiner}, {Yu}, {Zarrouk}, {Zhang}, {Zou}, {Ahlen}, {Bailey}, {Brooks}, {Cooper}, {de la Macorra}, {Dey}, {Dhungana}, {Doel}, {Eftekharzadeh}, {Fanning}, {Font-Ribera}, {Garrison}, {Gazta{\~n}aga}, {Kehoe}, {Kisner}, {Kremin}, {Landriau}, {Le Guillou}, {Levi}, {Magneville}, {Meisner}, {Miquel}, {Moustakas}, {Myers}, {Newman}, {Nie}, {Palanque-Delabrouille}, {Percival}, {Poppett}, {Prada}, {Schubnell}, {Tarl{\'e}}, {Weaver}, {Zhang}, \& {Zhou}}]{Lan2023}
{Lan}, T.-W., {Tojeiro}, R., {Armengaud}, E., {et~al.} 2023, \bibinfo{title}{{The DESI Survey Validation: Results from Visual Inspection of Bright Galaxies, Luminous Red Galaxies, and Emission-line Galaxies},} \apj, 943, 68, \dodoi{10.3847/1538-4357/aca5fa}

\bibitem[{D. {Lang} {et~al.}(2016){Lang}, {Hogg}, \& {Mykytyn}}]{Lang2016}
{Lang}, D., {Hogg}, D.~W., \& {Mykytyn}, D. 2016, \bibinfo{title}{{The Tractor: Probabilistic astronomical source detection and measurement},}, Astrophysics Source Code Library, record ascl:1604.008

\bibitem[{C.~A. {Lemon} {et~al.}(2019){Lemon}, {Auger}, \& {McMahon}}]{Lemon2019}
{Lemon}, C.~A., {Auger}, M.~W., \& {McMahon}, R.~G. 2019, \bibinfo{title}{{Gravitationally lensed quasars in Gaia - III. 22 new lensed quasars from Gaia data release 2},} \mnras, 483, 4242, \dodoi{10.1093/mnras/sty3366}

\bibitem[{C.~A. {Lemon} {et~al.}(2018){Lemon}, {Auger}, {McMahon}, \& {Ostrovski}}]{Lemon2018}
{Lemon}, C.~A., {Auger}, M.~W., {McMahon}, R.~G., \& {Ostrovski}, F. 2018, \bibinfo{title}{{Gravitationally lensed quasars in Gaia - II. Discovery of 24 lensed quasars},} \mnras, 479, 5060, \dodoi{10.1093/mnras/sty911}

\bibitem[{R. {Li} {et~al.}(2021){Li}, {Napolitano}, {Spiniello}, {Tortora}, {Kuijken}, {Koopmans}, {Schneider}, {Getman}, {Xie}, {Long}, {Shu}, {Vernardos}, {Huang}, {Covone}, {Dvornik}, {Heymans}, {Hildebrandt}, {Radovich}, \& {Wright}}]{Li2021}
{Li}, R., {Napolitano}, N.~R., {Spiniello}, C., {et~al.} 2021, \bibinfo{title}{{High-quality Strong Lens Candidates in the Final Kilo-Degree Survey Footprint},} \apj, 923, 16, \dodoi{10.3847/1538-4357/ac2df0}

\bibitem[{S. {Li} {et~al.}(2025){Li}, {Li}, {Wang}, {Jia}, {Cao}, {Frenk}, {Jiang}, {Amvrosiadis}, {Cole}, {He}, {Lange}, {Massey}, {Nightingale}, {Robertson}, {von Wietersheim-Kramsta}, \& {Ma}}]{Li2025}
{Li}, S., {Li}, R., {Wang}, K., {et~al.} 2025, \bibinfo{title}{{The ``Little Dark Dot'': Evidence for Self-Interacting Dark Matter in the Strong Lens SDSSJ0946+1006?},} arXiv e-prints, arXiv:2504.11800, \dodoi{10.48550/arXiv.2504.11800}

\bibitem[{A. {Manj{\'o}n-Garc{\'\i}a} {et~al.}(2019){Manj{\'o}n-Garc{\'\i}a}, {Herranz}, {Diego}, {Bonavera}, \& {Gonz{\'a}lez-Nuevo}}]{Manjon-Garcia2019}
{Manj{\'o}n-Garc{\'\i}a}, A., {Herranz}, D., {Diego}, J.~M., {Bonavera}, L., \& {Gonz{\'a}lez-Nuevo}, J. 2019, \bibinfo{title}{{Multifrequency filter search for high redshift sources and lensing systems in Herschel-ATLAS},} \aap, 622, A106, \dodoi{10.1051/0004-6361/201834549}

\bibitem[{P.~J. {Marshall} {et~al.}(2016){Marshall}, {Verma}, {More}, {Davis}, {More}, {Kapadia}, {Parrish}, {Snyder}, {Wilcox}, {Baeten}, {Macmillan}, {Cornen}, {Baumer}, {Simpson}, {Lintott}, {Miller}, {Paget}, {Simpson}, {Smith}, {K{\"u}ng}, {Saha}, \& {Collett}}]{Marshall2016}
{Marshall}, P.~J., {Verma}, A., {More}, A., {et~al.} 2016, \bibinfo{title}{{SPACE WARPS - I. Crowdsourcing the discovery of gravitational lenses},} \mnras, 455, 1171, \dodoi{10.1093/mnras/stv2009}

\bibitem[{T.~N. {Miller} {et~al.}(2024){Miller}, {Doel}, {Gutierrez}, {Besuner}, {Brooks}, {Gallo}, {Heetderks}, {Jelinsky}, {Kent}, {Lampton}, {Levi}, {Liang}, {Meisner}, {Sholl}, {Silber}, {Sprayberry}, {Aguilar}, {de la Macorra}, {Eisenstein}, {Fanning}, {Font-Ribera}, {Gazta{\~n}aga}, {Gontcho A Gontcho}, {Honscheid}, {Jimenez}, {Joyce}, {Kehoe}, {Kisner}, {Kremin}, {Landriau}, {Le Guillou}, {Magneville}, {Martini}, {Miquel}, {Moustakas}, {Nie}, {Percival}, {Poppett}, {Prada}, {Rossi}, {Schlegel}, {Schubnell}, {Seo}, {Sharples}, {Tarl{\'e}}, {Vargas-Maga{\~n}a}, {Zhou}, \& {the DESI Collaboration}}]{Miller2024}
{Miller}, T.~N., {Doel}, P., {Gutierrez}, G., {et~al.} 2024, \bibinfo{title}{{The Optical Corrector for the Dark Energy Spectroscopic Instrument},} \aj, 168, 95, \dodoi{10.3847/1538-3881/ad45fe}

\bibitem[{A. {More} {et~al.}(2016{\natexlab{a}}){More}, {Verma}, {Marshall}, {More}, {Baeten}, {Wilcox}, {Macmillan}, {Cornen}, {Kapadia}, {Parrish}, {Snyder}, {Davis}, {Gavazzi}, {Lintott}, {Simpson}, {Miller}, {Smith}, {Paget}, {Saha}, {K{\"u}ng}, \& {Collett}}]{More2016b}
{More}, A., {Verma}, A., {Marshall}, P.~J., {et~al.} 2016{\natexlab{a}}, \bibinfo{title}{{SPACE WARPS- II. New gravitational lens candidates from the CFHTLS discovered through citizen science},} \mnras, 455, 1191, \dodoi{10.1093/mnras/stv1965}

\bibitem[{A. {More} {et~al.}(2016{\natexlab{b}}){More}, {Oguri}, {Kayo}, {Zinn}, {Strauss}, {Santiago}, {Mosquera}, {Inada}, {Kochanek}, {Rusu}, {Brownstein}, {da Costa}, {Kneib}, {Maia}, {Quimby}, {Schneider}, {Streblyanska}, \& {York}}]{More2016a}
{More}, A., {Oguri}, M., {Kayo}, I., {et~al.} 2016{\natexlab{b}}, \bibinfo{title}{{The SDSS-III BOSS quasar lens survey: discovery of 13 gravitationally lensed quasars},} \mnras, 456, 1595, \dodoi{10.1093/mnras/stv2813}

\bibitem[{J. {Moustakas} {et~al.}(2023){Moustakas}, {Buhler}, {Scholte}, {Dey}, \& {Khederlarian}}]{Moustakas2023}
{Moustakas}, J., {Buhler}, J., {Scholte}, D., {Dey}, B., \& {Khederlarian}, A. 2023, \bibinfo{title}{{FastSpecFit: Fast spectral synthesis and emission-line fitting of DESI spectra},}, Astrophysics Source Code Library, record ascl:2308.005

\bibitem[{A.~D. {Myers} {et~al.}(2023){Myers}, {Moustakas}, {Bailey}, {Weaver}, {Cooper}, {Forero-Romero}, {Abolfathi}, {Alexander}, {Brooks}, {Chaussidon}, {Chuang}, {Dawson}, {Dey}, {Dey}, {Dhungana}, {Doel}, {Fanning}, {Gazta{\~n}aga}, {Gontcho A Gontcho}, {Gonzalez-Morales}, {Hahn}, {Herrera-Alcantar}, {Honscheid}, {Ishak}, {Karim}, {Kirkby}, {Kisner}, {Koposov}, {Kremin}, {Lan}, {Landriau}, {Lang}, {Levi}, {Magneville}, {Napolitano}, {Martini}, {Meisner}, {Newman}, {Palanque-Delabrouille}, {Percival}, {Poppett}, {Prada}, {Raichoor}, {Ross}, {Schlafly}, {Schlegel}, {Schubnell}, {Tan}, {Tarle}, {Wilson}, {Y{\`e}che}, {Zhou}, {Zhou}, \& {Zou}}]{Myers2023}
{Myers}, A.~D., {Moustakas}, J., {Bailey}, S., {et~al.} 2023, \bibinfo{title}{{The Target-selection Pipeline for the Dark Energy Spectroscopic Instrument},} \aj, 165, 50, \dodoi{10.3847/1538-3881/aca5f9}

\bibitem[{R. {Narayan} \& M. {Bartelmann}(1996){Narayan} \& {Bartelmann}}]{Narayan1996}
{Narayan}, R., \& {Bartelmann}, M. 1996, \bibinfo{title}{{Lectures on Gravitational Lensing},} arXiv e-prints, astro, \dodoi{10.48550/arXiv.astro-ph/9606001}

\bibitem[{A.~B. {Newman} {et~al.}(2015){Newman}, {Ellis}, \& {Treu}}]{Newman2015}
{Newman}, A.~B., {Ellis}, R.~S., \& {Treu}, T. 2015, \bibinfo{title}{{Luminous and Dark Matter Profiles from Galaxies to Clusters: Bridging the Gap with Group-scale Lenses},} \apj, 814, 26, \dodoi{10.1088/0004-637X/814/1/26}

\bibitem[{J.~H. {O'Donnell} {et~al.}(2022){O'Donnell}, {Wilkinson}, {Diehl}, {Aros-Bunster}, {Bechtol}, {Birrer}, {Buckley-Geer}, {Carnero Rosell}, {Carrasco Kind}, {da Costa}, {Gonzalez Lozano}, {Gruendl}, {Hilton}, {Lin}, {Lindgren}, {Martin}, {Pieres}, {Rykoff}, {Sevilla-Noarbe}, {Sheldon}, {Sif{\'o}n}, {Tucker}, {Yanny}, {Abbott}, {Aguena}, {Allam}, {Andrade-Oliveira}, {Annis}, {Bertin}, {Brooks}, {Burke}, {Carretero}, {Costanzi}, {De Vicente}, {Desai}, {Dietrich}, {Eckert}, {Everett}, {Ferrero}, {Flaugher}, {Fosalba}, {Frieman}, {Garc{\'\i}a-Bellido}, {Gaztanaga}, {Gerdes}, {Gruen}, {Gschwend}, {Gill}, {Gutierrez}, {Hinton}, {Hollowood}, {Honscheid}, {James}, {Jeltema}, {Kuehn}, {Lahav}, {Lima}, {Maia}, {Marshall}, {Melchior}, {Menanteau}, {Miquel}, {Morgan}, {Nord}, {Ogando}, {Paz-Chinch{\'o}n}, {Pereira}, {Plazas Malag{\'o}n}, {Rodriguez-Monroy}, {Romer}, {Roodman}, {Sanchez}, {Scarpine}, {Schubnell}, {Serrano}, {Smith}, {Suchyta}, {Swanson}, {Tarle}, {Thomas}, {To}, \& {Varga}}]{O'Donnell2022}
{O'Donnell}, J.~H., {Wilkinson}, R.~D., {Diehl}, H.~T., {et~al.} 2022, \bibinfo{title}{{The Dark Energy Survey Bright Arcs Survey: Candidate Strongly Lensed Galaxy Systems from the Dark Energy Survey 5000 Square Degree Footprint},} \apjs, 259, 27, \dodoi{10.3847/1538-4365/ac470b}

\bibitem[{C.~E. {Petrillo} {et~al.}(2019){Petrillo}, {Tortora}, {Vernardos}, {Koopmans}, {Verdoes Kleijn}, {Bilicki}, {Napolitano}, {Chatterjee}, {Covone}, {Dvornik}, {Erben}, {Getman}, {Giblin}, {Heymans}, {de Jong}, {Kuijken}, {Schneider}, {Shan}, {Spiniello}, \& {Wright}}]{Petrillo2019}
{Petrillo}, C.~E., {Tortora}, C., {Vernardos}, G., {et~al.} 2019, \bibinfo{title}{{LinKS: discovering galaxy-scale strong lenses in the Kilo-Degree Survey using convolutional neural networks},} \mnras, 484, 3879, \dodoi{10.1093/mnras/stz189}

\bibitem[{C. {Poppett} {et~al.}(2024){Poppett}, {Tyas}, {Aguilar}, {Bebek}, {Bramall}, {Claybaugh}, {Edelstein}, {Fagrelius}, {Heetderks}, {Jelinsky}, {Jelinsky}, {Lafever}, {Lambert}, {Lampton}, {Levi}, {Martini}, {Rockosi}, {Schmoll}, {Sharples}, {Sirk}, {Wishnow}, {Yu}, {Ahlen}, {Bault}, {BenZvi}, {Brooks}, {Cole}, {de la Macorra}, {Dey}, {Doel}, {Fanning}, {Font-Ribera}, {Forero-Romero}, {Gazta{\~n}aga}, {Gontcho A Gontcho}, {Gonzalez-Morales}, {Hahn}, {Honscheid}, {Jimenez}, {Juneau}, {Kirkby}, {Kremin}, {Landriau}, {Le Guillou}, {Manera}, {Meisner}, {Miquel}, {Moustakas}, {Mueller}, {Mu{\~n}oz-Guti{\'e}rrez}, {Myers}, {Nie}, {Niz}, {Palanque-Delabrouille}, {Percival}, {Prada}, {Rabinowitz}, {Rezaie}, {Rossi}, {Sanchez}, {Schlafly}, {Schlegel}, {Schubnell}, {Seo}, {Sprayberry}, {Tarl{\'e}}, {Vargas-Maga{\~n}a}, {Weaver}, \& {Zhou}}]{Poppett2024}
{Poppett}, C., {Tyas}, L., {Aguilar}, J., {et~al.} 2024, \bibinfo{title}{{Overview of the Fiber System for the Dark Energy Spectroscopic Instrument},} \aj, 168, 245, \dodoi{10.3847/1538-3881/ad76a4}

\bibitem[{A. {Raichoor} {et~al.}(2023){Raichoor}, {Moustakas}, {Newman}, {Karim}, {Ahlen}, {Alam}, {Bailey}, {Brooks}, {Dawson}, {de la Macorra}, {de Mattia}, {Dey}, {Dey}, {Dhungana}, {Eftekharzadeh}, {Eisenstein}, {Fanning}, {Font-Ribera}, {Garc{\'\i}a-Bellido}, {Gazta{\~n}aga}, {A Gontcho}, {Guy}, {Honscheid}, {Ishak}, {Kehoe}, {Kisner}, {Kremin}, {Lan}, {Landriau}, {Le Guillou}, {Levi}, {Magneville}, {Manera}, {Martini}, {Meisner}, {Myers}, {Nie}, {Palanque-Delabrouille}, {Percival}, {Poppett}, {Prada}, {Ross}, {Ruhlmann-Kleider}, {Sabiu}, {Schlafly}, {Schlegel}, {Tarl{\'e}}, {Weaver}, {Y{\`e}che}, {Zhou}, {Zhou}, \& {Zou}}]{Raichoor2023}
{Raichoor}, A., {Moustakas}, J., {Newman}, J.~A., {et~al.} 2023, \bibinfo{title}{{Target Selection and Validation of DESI Emission Line Galaxies},} \aj, 165, 126, \dodoi{10.3847/1538-3881/acb213}

\bibitem[{J.~I. {Read} {et~al.}(2017){Read}, {Iorio}, {Agertz}, \& {Fraternali}}]{Read2017}
{Read}, J.~I., {Iorio}, G., {Agertz}, O., \& {Fraternali}, F. 2017, \bibinfo{title}{{The stellar mass-halo mass relation of isolated field dwarfs: a critical test of {\ensuremath{\Lambda}}CDM at the edge of galaxy formation},} \mnras, 467, 2019, \dodoi{10.1093/mnras/stx147}

\bibitem[{G. {Roberts-Borsani} {et~al.}(2022){Roberts-Borsani}, {Morishita}, {Treu}, {Brammer}, {Strait}, {Wang}, {Bradac}, {Acebron}, {Bergamini}, {Boyett}, {Calabr{\'o}}, {Castellano}, {Fontana}, {Glazebrook}, {Grillo}, {Henry}, {Jones}, {Malkan}, {Marchesini}, {Mascia}, {Mason}, {Mercurio}, {Merlin}, {Nanayakkara}, {Pentericci}, {Rosati}, {Santini}, {Scarlata}, {Trenti}, {Vanzella}, {Vulcani}, \& {Willott}}]{Roberts-Borsani2022}
{Roberts-Borsani}, G., {Morishita}, T., {Treu}, T., {et~al.} 2022, \bibinfo{title}{{Early Results from GLASS-JWST. I: Confirmation of Lensed z {\ensuremath{\geq}} 7 Lyman-break Galaxies behind the Abell 2744 Cluster with NIRISS},} \apjl, 938, L13, \dodoi{10.3847/2041-8213/ac8e6e}

\bibitem[{E. {Savary} {et~al.}(2022){Savary}, {Rojas}, {Maus}, {Cl{\'e}ment}, {Courbin}, {Gavazzi}, {Chan}, {Lemon}, {Vernardos}, {Ca{\~n}ameras}, {Schuldt}, {Suyu}, {Cuillandre}, {Fabbro}, {Gwyn}, {Hudson}, {Kilbinger}, {Scott}, \& {Stone}}]{Savary2022}
{Savary}, E., {Rojas}, K., {Maus}, M., {et~al.} 2022, \bibinfo{title}{{Strong lensing in UNIONS: Toward a pipeline from discovery to modeling},} \aap, 666, A1, \dodoi{10.1051/0004-6361/202142505}

\bibitem[{E.~F. {Schlafly} {et~al.}(2023){Schlafly}, {Kirkby}, {Schlegel}, {Myers}, {Raichoor}, {Dawson}, {Aguilar}, {Allende Prieto}, {Bailey}, {BenZvi}, {Bermejo-Climent}, {Brooks}, {de la Macorra}, {Dey}, {Doel}, {Fanning}, {Font-Ribera}, {Forero-Romero}, {Garc{\'\i}a-Bellido}, {Gontcho A Gontcho}, {Guy}, {Hahn}, {Honscheid}, {Ishak}, {Juneau}, {Kehoe}, {Kisner}, {Kremin}, {Landriau}, {Lang}, {Lasker}, {Levi}, {Magneville}, {Manser}, {Martini}, {Meisner}, {Miquel}, {Moustakas}, {Newman}, {Nie}, {Palanque-Delabrouille}, {Percival}, {Poppett}, {Rockosi}, {Ross}, {Rossi}, {Tarl{\'e}}, {Weaver}, {Y{\`e}che}, {Zhou}, \& {DESI Collaboration}}]{Schlafly2023}
{Schlafly}, E.~F., {Kirkby}, D., {Schlegel}, D.~J., {et~al.} 2023, \bibinfo{title}{{Survey Operations for the Dark Energy Spectroscopic Instrument},} \aj, 166, 259, \dodoi{10.3847/1538-3881/ad0832}

\bibitem[{Y. {Shu} {et~al.}(2022){Shu}, {Ca{\~n}ameras}, {Schuldt}, {Suyu}, {Taubenberger}, {Inoue}, \& {Jaelani}}]{Shu2022}
{Shu}, Y., {Ca{\~n}ameras}, R., {Schuldt}, S., {et~al.} 2022, \bibinfo{title}{{HOLISMOKES. VIII. High-redshift, strong-lens search in the Hyper Suprime-Cam Subaru Strategic Program},} \aap, 662, A4, \dodoi{10.1051/0004-6361/202243203}

\bibitem[{J.~H. {Silber} {et~al.}(2023){Silber}, {Fagrelius}, {Fanning}, {Schubnell}, {Aguilar}, {Ahlen}, {Ameel}, {Ballester}, {Baltay}, {Bebek}, \& et~al.}]{Silber2023}
{Silber}, J.~H., {Fagrelius}, P., {Fanning}, K., {et~al.} 2023, \bibinfo{title}{{The Robotic Multiobject Focal Plane System of the Dark Energy Spectroscopic Instrument (DESI)},} \aj, 165, 9, \dodoi{10.3847/1538-3881/ac9ab1}

\bibitem[{E. {Silver} {et~al.}(2025){Silver}, {Wang}, {Huang}, {Bolton}, {Storfer}, \& {Banka}}]{Silver2025}
{Silver}, E., {Wang}, R., {Huang}, X., {et~al.} 2025, \bibinfo{title}{{ML-Driven Strong Lens Discoveries: Down to $θ_E \sim 0.03''$ and $M_\mathrm{halo}< 10^{11} M_\odot$},} arXiv e-prints, arXiv:2507.01943, \dodoi{10.48550/arXiv.2507.01943}

\bibitem[{A. {Sonnenfeld} {et~al.}(2013){Sonnenfeld}, {Gavazzi}, {Suyu}, {Treu}, \& {Marshall}}]{Sonnenfeld2013}
{Sonnenfeld}, A., {Gavazzi}, R., {Suyu}, S.~H., {Treu}, T., \& {Marshall}, P.~J. 2013, \bibinfo{title}{{The SL2S Galaxy-scale Lens Sample. III. Lens Models, Surface Photometry, and Stellar Masses for the Final Sample},} \apj, 777, 97, \dodoi{10.1088/0004-637X/777/2/97}

\bibitem[{A. {Sonnenfeld} {et~al.}(2019){Sonnenfeld}, {Jaelani}, {Chan}, {More}, {Suyu}, {Wong}, {Oguri}, \& {Lee}}]{Sonnenfeld2019}
{Sonnenfeld}, A., {Jaelani}, A.~T., {Chan}, J., {et~al.} 2019, \bibinfo{title}{{Survey of gravitationally-lensed objects in HSC imaging (SuGOHI). III. Statistical strong lensing constraints on the stellar IMF of CMASS galaxies},} \aap, 630, A71, \dodoi{10.1051/0004-6361/201935743}

\bibitem[{A. {Sonnenfeld} {et~al.}(2018){Sonnenfeld}, {Chan}, {Shu}, {More}, {Oguri}, {Suyu}, {Wong}, {Lee}, {Coupon}, {Yonehara}, {Bolton}, {Jaelani}, {Tanaka}, {Miyazaki}, \& {Komiyama}}]{Sonnenfeld2018}
{Sonnenfeld}, A., {Chan}, J. H.~H., {Shu}, Y., {et~al.} 2018, \bibinfo{title}{{Survey of Gravitationally-lensed Objects in HSC Imaging (SuGOHI). I. Automatic search for galaxy-scale strong lenses},} \pasj, 70, S29, \dodoi{10.1093/pasj/psx062}

\bibitem[{A. {Sonnenfeld} {et~al.}(2020){Sonnenfeld}, {Verma}, {More}, {Baeten}, {Macmillan}, {Wong}, {Chan}, {Jaelani}, {Lee}, {Oguri}, {Rusu}, {Veldthuis}, {Trouille}, {Marshall}, {Hutchings}, {Allen}, {O'Donnell}, {Cornen}, {Davis}, {McMaster}, {Lintott}, \& {Miller}}]{Sonnenfeld2020}
{Sonnenfeld}, A., {Verma}, A., {More}, A., {et~al.} 2020, \bibinfo{title}{{Survey of Gravitationally-lensed Objects in HSC Imaging (SuGOHI). VI. Crowdsourced lens finding with Space Warps},} \aap, 642, A148, \dodoi{10.1051/0004-6361/202038067}

\bibitem[{V. {Springel} {et~al.}(2008){Springel}, {Wang}, {Vogelsberger}, {Ludlow}, {Jenkins}, {Helmi}, {Navarro}, {Frenk}, \& {White}}]{Springel2008}
{Springel}, V., {Wang}, J., {Vogelsberger}, M., {et~al.} 2008, \bibinfo{title}{{The Aquarius Project: the subhaloes of galactic haloes},} \mnras, 391, 1685, \dodoi{10.1111/j.1365-2966.2008.14066.x}

\bibitem[{G. {Stein} {et~al.}(2022){Stein}, {Blaum}, {Harrington}, {Medan}, \& {Luki{\'c}}}]{Stein2022}
{Stein}, G., {Blaum}, J., {Harrington}, P., {Medan}, T., \& {Luki{\'c}}, Z. 2022, \bibinfo{title}{{Mining for Strong Gravitational Lenses with Self-supervised Learning},} \apj, 932, 107, \dodoi{10.3847/1538-4357/ac6d63}

\bibitem[{C. {Storfer} {et~al.}(2024){Storfer}, {Huang}, {Gu}, {Sheu}, {Banka}, {Dey}, {Inchausti Reyes}, {Jain}, {Kwon}, {Lang}, {Lee}, {Meisner}, {Moustakas}, {Myers}, {Tabares-Tarquinio}, {Schlafly}, \& {Schlegel}}]{Storfer2024}
{Storfer}, C., {Huang}, X., {Gu}, A., {et~al.} 2024, \bibinfo{title}{{New Strong Gravitational Lenses from the DESI Legacy Imaging Surveys Data Release 9},} \apjs, 274, 16, \dodoi{10.3847/1538-4365/ad527e}

\bibitem[{C.~J. {Storfer} {et~al.}(2025){Storfer}, {Magnier}, {Huang}, {Rubin}, {Schlegel}, {Banka}, {Chambers}, {Cuillandre}, {de Boer}, {Gavazzi}, {Gwyn}, {Hudson}, {Paek}, \& {Scott}}]{Storfer2025}
{Storfer}, C.~J., {Magnier}, E.~A., {Huang}, X., {et~al.} 2025, \bibinfo{title}{{Gravitational Lenses in UNIONS and Euclid (GLUE) I: A Search for Strong Gravitational Lenses in UNIONS with Subaru, CFHT, and Pan-STARRS Data},} arXiv e-prints, arXiv:2505.05032, \dodoi{10.48550/arXiv.2505.05032}

\bibitem[{M.~S. {Talbot} {et~al.}(2021){Talbot}, {Brownstein}, {Dawson}, {Kneib}, \& {Bautista}}]{Talbot2021}
{Talbot}, M.~S., {Brownstein}, J.~R., {Dawson}, K.~S., {Kneib}, J.-P., \& {Bautista}, J. 2021, \bibinfo{title}{{The completed SDSS-IV extended Baryon Oscillation Spectroscopic Survey: a catalogue of strong galaxy-galaxy lens candidates},} \mnras, 502, 4617, \dodoi{10.1093/mnras/stab267}

\bibitem[{ {TDCOSMO Collaboration} {et~al.}(2025){TDCOSMO Collaboration}, {Birrer}, {Buckley-Geer}, {Cappellari}, {Courbin}, {Dux}, {Fassnacht}, {Frieman}, {Galan}, {Gilman}, {Huang}, {Knabel}, {Langeroodi}, {Lin}, {Millon}, {Morishita}, {Motta}, {Mozumdar}, {Paic}, {Shajib}, {Sheu}, {Sluse}, {Sonnenfeld}, {Spiniello}, {Stiavelli}, {Suyu}, {Tan}, {Treu}, {Van de Vyvere}, {Wang}, {Wells}, {Williams}, \& {Wong}}]{TDCOSMOCollaboration2025}
{TDCOSMO Collaboration}, {Birrer}, S., {Buckley-Geer}, E.~J., {et~al.} 2025, \bibinfo{title}{{TDCOSMO 2025: Cosmological constraints from strong lensing time delays},} arXiv e-prints, arXiv:2506.03023, \dodoi{10.48550/arXiv.2506.03023}

\bibitem[{S. {Vegetti} {et~al.}(2010){Vegetti}, {Koopmans}, {Bolton}, {Treu}, \& {Gavazzi}}]{Vegetti2010}
{Vegetti}, S., {Koopmans}, L.~V.~E., {Bolton}, A., {Treu}, T., \& {Gavazzi}, R. 2010, \bibinfo{title}{{Detection of a dark substructure through gravitational imaging},} \mnras, 408, 1969, \dodoi{10.1111/j.1365-2966.2010.16865.x}

\bibitem[{M.~S. {Warren} {et~al.}(2006){Warren}, {Abazajian}, {Holz}, \& {Teodoro}}]{Warren2006}
{Warren}, M.~S., {Abazajian}, K., {Holz}, D.~E., \& {Teodoro}, L. 2006, \bibinfo{title}{{Precision Determination of the Mass Function of Dark Matter Halos},} \apj, 646, 881, \dodoi{10.1086/504962}

\bibitem[{R.~H. {Wechsler} \& J.~L. {Tinker}(2018){Wechsler} \& {Tinker}}]{Wechsler2018}
{Wechsler}, R.~H., \& {Tinker}, J.~L. 2018, \bibinfo{title}{{The Connection Between Galaxies and Their Dark Matter Halos},} \araa, 56, 435, \dodoi{10.1146/annurev-astro-081817-051756}

\bibitem[{K.~C. {Wong} {et~al.}(2022){Wong}, {Chan}, {Chao}, {Jaelani}, {Kayo}, {Lee}, {More}, \& {Oguri}}]{Wong2022}
{Wong}, K.~C., {Chan}, J. H.~H., {Chao}, D. C.~Y., {et~al.} 2022, \bibinfo{title}{{Survey of Gravitationally lensed objects in HSC Imaging (SuGOHI). VIII. New galaxy-scale lenses from the HSC SSP},} \pasj, 74, 1209, \dodoi{10.1093/pasj/psac065}

\bibitem[{K.~C. {Wong} {et~al.}(2018){Wong}, {Sonnenfeld}, {Chan}, {Rusu}, {Tanaka}, {Jaelani}, {Lee}, {More}, {Oguri}, {Suyu}, \& {Komiyama}}]{Wong2018}
{Wong}, K.~C., {Sonnenfeld}, A., {Chan}, J. H.~H., {et~al.} 2018, \bibinfo{title}{{Survey of Gravitationally Lensed Objects in HSC Imaging (SuGOHI). II. Environments and Line-of-Sight Structure of Strong Gravitational Lens Galaxies to z {\ensuremath{\sim}} 0.8},} \apj, 867, 107, \dodoi{10.3847/1538-4357/aae381}

\bibitem[{G. {Zhang} {et~al.}(2024){Zhang}, {{\c{S}}eng{\"u}l}, \& {Dvorkin}}]{Zhang2024}
{Zhang}, G., {{\c{S}}eng{\"u}l}, A.~{\c{C}}., \& {Dvorkin}, C. 2024, \bibinfo{title}{{Subhalo effective density slope measurements from HST strong lensing data with neural likelihood-ratio estimation},} \mnras, 527, 4183, \dodoi{10.1093/mnras/stad3521}

\bibitem[{R. {Zhou} {et~al.}(2023){Zhou}, {Dey}, {Newman}, {Eisenstein}, {Dawson}, {Bailey}, {Berti}, {Guy}, {Lan}, {Zou}, {Aguilar}, {Ahlen}, {Alam}, {Brooks}, {de la Macorra}, {Dey}, {Dhungana}, {Fanning}, {Font-Ribera}, {Gontcho}, {Honscheid}, {Ishak}, {Kisner}, {Kov{\'a}cs}, {Kremin}, {Landriau}, {Levi}, {Magneville}, {Manera}, {Martini}, {Meisner}, {Miquel}, {Moustakas}, {Myers}, {Nie}, {Palanque-Delabrouille}, {Percival}, {Poppett}, {Prada}, {Raichoor}, {Ross}, {Schlafly}, {Schlegel}, {Schubnell}, {Tarl{\'e}}, {Weaver}, {Wechsler}, {Y{\'e}che}, \& {Zhou}}]{Zhou2023}
{Zhou}, R., {Dey}, B., {Newman}, J.~A., {et~al.} 2023, \bibinfo{title}{{Target Selection and Validation of DESI Luminous Red Galaxies},} \aj, 165, 58, \dodoi{10.3847/1538-3881/aca5fb}

\end{thebibliography}
\bibliographystyle{aasjournalv7}

\end{CJK*}
\end{document}